\documentclass[acmsmall]{acmart}
\usepackage{lscape}
\usepackage{geometry}
\newcommand{\leaveout}[1]{}
\newcommand{\yx}[1]{{\color{black} #1}}
\newcommand{\zz}[1]{{\color{black} #1}}

\AtBeginDocument{%
  }

\setcopyright{acmcopyright}
\copyrightyear{2023}
\acmYear{2023}
\acmDOI{XXXXXXX.XXXXXXX}

\acmVolume{00}
\acmNumber{0}
\acmArticle{000}
\acmMonth{0}

\begin{document}

\title{Trusta: Reasoning about Assurance Cases with Formal Methods and Large Language Models}





\author{Zezhong Chen}
\affiliation{%
  \institution{Shanghai Key Laboratory of Trustworthy Computing, East China Normal University}
   \streetaddress{3663 Zhongshan North Road}
  \city{Shanghai}
  \country{China}
  \postcode{200062}
}

\author{Yuxin Deng}
\email{yxdeng@sei.ecnu.edu.cn}
\affiliation{%
\institution{Shanghai Key Laboratory of Trustworthy Computing, East China Normal University}
  \streetaddress{3663 Zhongshan North Road}
\city{Shanghai}
\country{China}
\postcode{200062}
}

 \author{Wenjie Du}
 \affiliation{%
   \institution{Shanghai Normal University}
   \streetaddress{100 Guilin Road}
   \city{Shanghai}
   \country{China}
   \postcode{200233}}



\renewcommand{\shortauthors}{Z. Chen,  Y. Deng and W. Du}

\begin{abstract}
Assurance cases can be used to argue for the safety of products in safety engineering. In safety-critical areas, the construction of assurance cases is indispensable.
Trustworthiness Derivation Trees (TDTs) \zz{enhance} assurance cases by incorporating formal methods, rendering it possible for automatic reasoning about assurance cases. We present Trustworthiness Derivation Tree Analyzer (Trusta), a desktop application designed to automatically construct and verify TDTs.
The tool has a built-in Prolog interpreter in its backend, and is supported by the constraint solvers Z3 and MONA. Therefore, it can solve constraints about logical formulas involving arithmetic, sets, Horn clauses etc. 
\yx{Trusta also utilizes large language models to make the creation and evaluation of assurance cases more convenient. It allows for interactive human examination and modification.}
\zz{We evaluated top language models like ChatGPT-3.5, ChatGPT-4, and PaLM 2 for generating assurance cases. Our tests showed a 50\%-80\% similarity between machine-generated and human-created cases.}
\zz{\yx{In addition,} Trusta can \yx{extract formal constraints from text in } natural languages, facilitating an easier interpretation and validation process. This extraction is subject to human review and correction, blending the best of automated efficiency with human insight.}
To our knowledge, \zz{this marks} the first \zz{integration of large language models in} automatic \zz{creating and} reasoning about assurance cases\zz{, bringing a novel approach to a traditional challenge.}
\zz{Through} several industrial case studies, Trusta has \zz{proven} to quickly find some subtle issues that are \zz{typically missed in} manual inspection, \zz{demonstrating its practical value in enhancing the assurance case development process.}
\end{abstract}

\ccsdesc[500]{Do Not Use This Code~Generate the Correct Terms for Your Paper}
\ccsdesc[300]{Do Not Use This Code~Generate the Correct Terms for Your Paper}
\ccsdesc{Do Not Use This Code~Generate the Correct Terms for Your Paper}
\ccsdesc[100]{Do Not Use This Code~Generate the Correct Terms for Your Paper}

\keywords{Assurance cases, trustworthiness derivation trees, \zz{large language models}, constraint solving
}

\received{20 February 2007}
\received[revised]{12 March 2009}
\received[accepted]{5 June 2009}

\maketitle

\section{Introduction}

In safety-critical areas such as medical, automotive, and avionics domains, the long-standing practice has showed that applying assurance cases~\cite{Bishop1998A,Bloomfield2009Safety,International2011Systems} can bring system reliability and safety to conform to relevant industrial standards.
An
assurance case is a documented body of evidence that provides
a valid argument so that a specified set of claims regarding
a product's properties are adequately justified for a given
application in a given environment. It can be graphically depicted as a finite tree whose 
root node represents the main claim about a system under consideration, and the leaf nodes stand for evidences.
The other nodes are composed of sub-claims and auxiliary components.
These sub-claims 
provide compelling, comprehensible, and valid cases~\cite{Sujan2016Should}. 
Assurance cases can demonstrate acceptable safety for a given system.
They 
prove to be useful for risk management. On one hand, they demonstrate that the risks associated with a system have been identified. On the other hand, they show that the risk mitigation measures have been effectively taken to ensure the system's safety performance. 
The assurance cases can also be a communication tool to bring different stakeholders to an agreement on the properties that should be satisfied by the system.

There exist a number of international functional safety standards that provide development guidelines for safety-critical systems such as ISO 26262~\cite{International2011Road} and DO-178C~\cite{RTCA2011Software}. In particular, the standard ISO 26262 explicitly recommends safety cases or assurance cases to demonstrate the safety of systems in the automotive domain.
Nowadays, assurance cases are widely used in the nuclear industry, the health and defense sectors, the oil industry, rail transport, automobile, and avionics~\cite{Rinehart2015Current,Rinehart2017Understanding}.
It is envisaged that they can be helpful in other areas, such as finance and telecommunications, which provide basic infrastructures for the whole society.
There exists a huge amount of literature arguing for a robust evidence-based approach for guaranteeing trustworthiness in software systems~\cite{National2007Software}, but most of the work on concrete assurance cases is not published due to various reasons such as security, confidentiality, and sensitivity.

Assurance cases for complex systems can be very large. For example, a typical assurance case for an air traffic control system may result in a document with over 500 pages and 400
referenced documents~\cite{Lewis2009Safety}.
The construction and evaluation of assurance cases is time-consuming as it 
 requires too much manual work.
As one of the steps in the overall safety certification process, a dedicated safety assessor is required to review and challenge the content of an assurance case.
During the evaluation process of an assurance case, the safety assessor is asked to evaluate the validity of the assurance case and discuss their judgment with the assurance case developers.
The high manual workload involved in
the construction and evaluation of assurance cases makes this process long and time-consuming.
The main challenge for the safety assessor is to check the loopholes in a large assurance case without omission.
To make things worse, the content of assurance cases is usually based on text description (informal description in natural languages), which may be ambiguous and is not amenable to automated assessment. Since the evaluation of assurance cases largely depends on human insight and experience, it is error prone due to faults in human judgment.
\zz{This complexity reveals the potential need for automation and artificial intelligence intervention, a gap that the introduction of the Trusta framework in this paper aims to address by combining large language models and human interaction in a novel and efficient way to create and reason about assurance cases.}

\zz{The need for automation in assurance case generation stems from the inherent complexity and resource-intensive nature of manually creating, maintaining, and updating assurance cases. Traditional methods often require significant expert involvement, extensive documentation, and meticulous tracking of claims, evidence, and arguments. This manual process can be error-prone, leading to potential inconsistencies and gaps that may jeopardize the integrity of the assurance case. Furthermore, as systems evolve and regulatory requirements change, updating assurance cases can become a cumbersome and time-consuming task. Automation offers the promise of efficiency, consistency, and adaptability, allowing for the real-time generation and updating of assurance cases, tailored to specific contexts and standards. The introduction of tools like Trusta that leverage advanced technologies such as large language models holds the potential to revolutionize the field by facilitating a more streamlined and dynamic approach to assurance case generation, thereby reducing the burden on human experts and enhancing overall effectiveness and reliability.}

\begin{figure}[!t]
\centering
\begin{minipage}[c]{0.54\textwidth}
\centering
\includegraphics[width=1\textwidth]{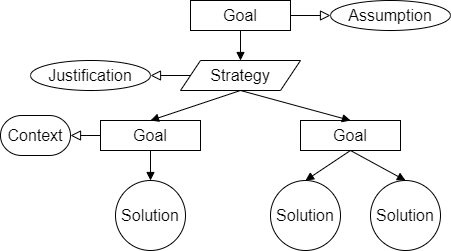}
\end{minipage}
\hspace{0.02\textwidth}
\begin{minipage}[c]{0.4\textwidth}
\centering
\includegraphics[width=1\textwidth]{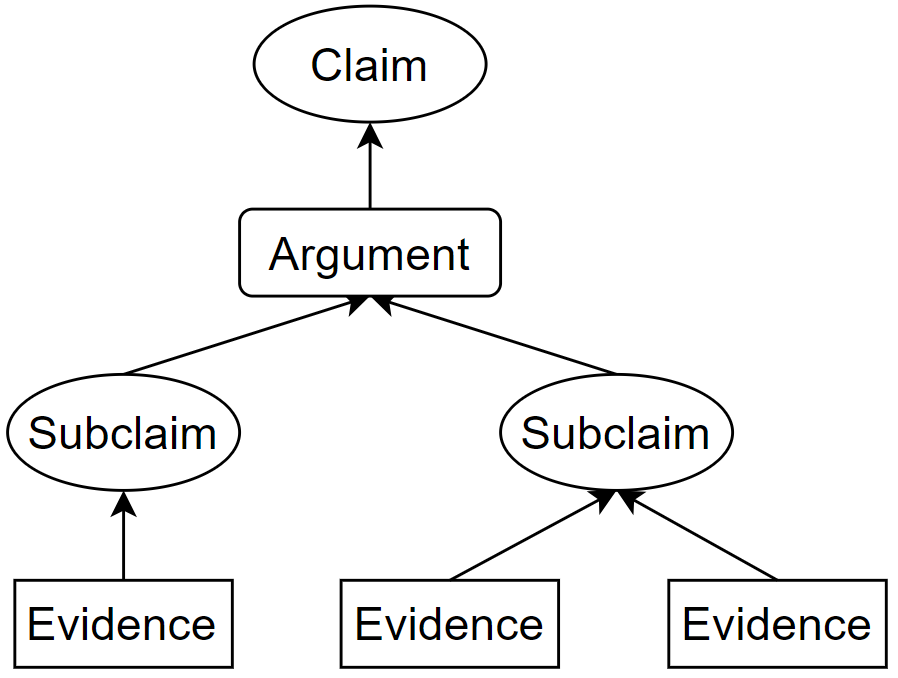}
\end{minipage}\\[3mm]
\begin{minipage}[t]{0.54\textwidth}
\centering
\caption{GSN notation.}
\label{demo_gsn}
\end{minipage}
\hspace{0.02\textwidth}
\begin{minipage}[t]{0.4\textwidth}
\centering
\caption{CAE notation.}
\label{demo_cae}
\end{minipage}
\end{figure}
\begin{figure}[!t]
\centering
\includegraphics[width=0.6\textwidth]{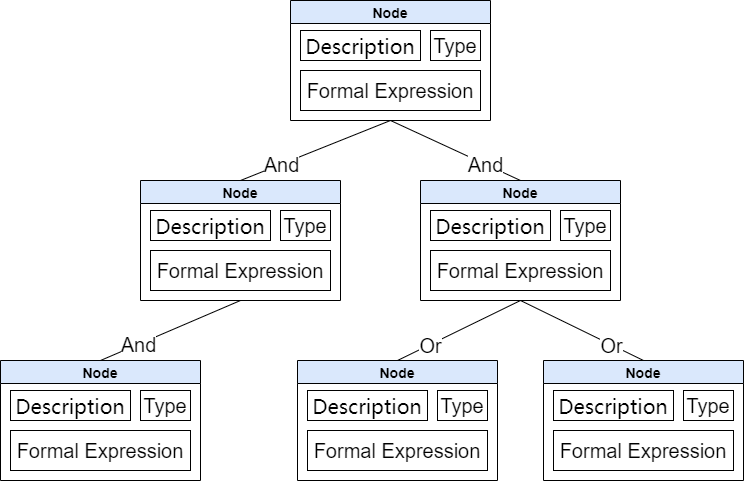}
\caption{\zz{TDT notation.}}
\label{demo_tdt}
\end{figure}

In order to facilitate the reasoning about assurance cases, we introduced the model of Trustworthiness Derivation Trees (TDTs)~\cite{Deng2021Trustworthiness} and exploited a few formal methods. A TDT is like an assurance case with only claims and evidences. An assurance case can be converted into a TDT in two steps:
(i) \zz{For assurance cases in the Goal Structuring Notation (GSN)~\cite{Weaver2004The, 2021The} format, turn the auxiliary components (contexts, assumptions, justifications, and strategies) into descriptions of nodes, while retaining the principal components (goals and solutions); for the Claim-Argument-Evidence (CAE)~\cite{Netkachova2014Tool} notation, the auxiliary components are represented by arguments and the principal components by claims and evidences;} 
(ii) Then add formal expressions and necessary parameters to express every principal component.
\zz{Appendix~\ref{appendix:CubeSat_TDT_GSN} showcases an example of the mutual conversion between an assurance case in GSN format and a TDT.}
By using formal expressions or logical formulas to represent the properties of a system, we open the door to automatic reasoning about TDTs and eventually about assurance cases.
Figure~\ref{demo_gsn} shows \zz{the widely recognized GSN} representation of assurance cases, \zz{while Figure~\ref{demo_cae} shows} the CAE notation. \zz{In contrast, Figure~\ref{demo_tdt} introduces our novel representation, the TDT. The unique aspect of the TDT, distinct from the GSN and CAE notations, is the incorporation of formal expressions. This} makes it possible to perform automatic reasoning from bottom to top.

In this paper, we introduce Trustworthiness Derivation Tree Analyzer (Trusta), which is a desktop application for automatically constructing and verifying TDTs. 
At the frontend, the tool provides a graphic user interface for creating and manipulating TDTs.
In its backend, a lightweight Prolog interpreter is built in. Moreover, it can invoke Z3~\cite{De2008Z3} and MONA~\cite{Klarlund2001Mona} to solve corresponding constraints in the formal expressions of \zz{goals.
Aided by a large language model, the backend also assists in breaking down a goal into sub-goals and helps in transforming natural language-formulated goals into constraint-based expressions.}
Currently, the allowed formal expressions are logical formulas involving arithmetic, sets, Horn clauses etc.
In a TDT, each \zz{node} is supported by \zz{several sub-node}s. The validity of the \zz{sub-nodes} implies \zz{the validity of the parent node}. Therefore, we can propagate the reasoning in a bottom-up fashion and eventually infer the validity of the root node of the tree.
We have conducted a few case studies such as  automated guided vehicles. Indeed, Trusta has helped us to quickly find some subtle problems that are otherwise difficult to spot by manual inspection. It also provides error analysis reports using the counterexamples output by the underlying constraint solvers.

Trusta simplifies assurance cases without losing their expressiveness, and is capable of performing automatic reasoning by incorporating formal methods.
It can help an assurance case developer to automatically identify potential errors and find the causes of the errors during the development process. 
Furthermore, it can help a safety assessor to find the errors that are difficult to detect manually.
The tool also provides a detailed report on which parts are at risk and what the risks are in a TDT. We believe that it can shorten the development cycle and improve safety for safety-critical systems.

\zz{There are two \yx{major} steps in the process of creating assurance cases with Trusta, both leveraging a large language model to assist the user in decision-making processes.
The first step involves the decomposition of a goal into sub-goals when creating nodes, a process that can be complex due to the nested nature and interconnected relationships within a goal. Trusta employs a language model to analyze the goal's structure and semantics, offering recommendations for suitable sub-goals that the user can then select or modify.
The second step is the formalization of the goal into a constraint formula, a task that demands precision and proper understanding of logical relations. Trusta's framework \yx{takes advantage of} the large language model's capability to comprehend and formulate mathematical and logical expressions, providing users with suggestions for converting the goal into a standardized constraint formula.
Both steps represent a fusion of machine intelligence with human oversight, aiming to alleviate some of the complexities and frustrations traditionally associated with assurance case generation, while still ensuring accuracy and flexibility through interactive user engagement.}

\zz{
In this \yx{article}, the application of a large language model serves as a critical innovation point within Trusta's assurance case generation process. By employing a series of specialized techniques, outlined in Section~\ref{sec:back}, we design prompt inputs that enable the language model to output structured information. Trusta's framework subsequently parses these outputs to present the required content either graphically or as mathematical expressions. More specifically, the application of the large language model unfolds in two key scenarios.
\begin{enumerate}
\item
Node creation in assurance cases: Within the input prompts, we incorporate not only theoretical knowledge concerning assurance cases but also the content of the current layer of assurance case nodes. The purpose of this approach is to enable the language model to generate meaningful content for subsequent layers. Trusta then parses these generated nodes and visually displays them, providing users with an intuitive means of understanding and modification.
\item
Conversion of \yx{text in a} natural language to constraint formulas: In this step, the input prompts are designed to encapsulate theoretical understanding of constraint-solving, along with the natural language expression awaiting transformation. The language model outputs the corresponding constraint-solving expression, which Trusta then parses into a standardized constraint formula. 
\end{enumerate}
}

\zz{
The main contributions of this \yx{article} can be summarized as follows:
\begin{enumerate}
\item Introduction of Trusta: A novel tool for enhancing assurance case creation through the integration of formal methods and large language models.
\item Intelligent automation: Trusta automates two of the most challenging steps in assurance case creation: the decomposition of goals into sub-goals and the translation of goals into constraint formulas, thereby providing smart recommendations.
\item Real-world applications and error analysis: \yx{We demonstrate} Trusta's practicality through case studies and its capability in identifying potential risks.
\item Cross-domain language model evaluation: A comprehensive study on the effectiveness of state-of-the-art language models (ChatGPT-3.5~\cite{OpenAI2023GPT35}, ChatGPT-4~\cite{OpenAI2023GPT4}, PaLM 2~\cite{Google2023PaLM2}) in generating assurance cases across multiple domains, revealing a 50\%-80\% similarity between machine-generated and human-created cases.
\end{enumerate}

By amalgamating human expertise with machine-driven insights, this article posits Trusta as a significant advancement in the field of safety-critical systems. Moreover, this research represents a major shift in the formal methods domain, offering a solution to the efficiency challenges commonly associated with the application of formal methods.
}

\zz{The remainder of this article is organized into distinct sections to provide a coherent and comprehensive overview of Trusta and its applications in assurance case generation and evaluation. Section~\ref{sec:back} delves into the theoretical background, elucidating the key concepts of assurance cases, large language models and constraint solvers. Section~\ref{sec:arch} introduces the architecture and functionalities of Trusta, with particular emphasis on the integration of large language models and their role in the two intricate steps of goal decomposition and goal translation. Section~\ref{sec:case} presents a case study that showcases the real-world application of Trusta in a safety-critical domain, followed by Section~\ref{sec:related} which offers a comparative analysis of Trusta with existing methodologies. \yx{F}inally, Section~\ref{sec:con} first concludes the paper by summarizing the key  contributions, and then discusses the future directions of the research. Appendices \yx{give a few concrete assurance cases to show the conversion between GSN and TDT formats.}
}
\section{Background}\label{sec:back}
\yx{In this section, we review some background knowledge about assurance cases, large language models, and constraint solvers.}
\subsection{Assurance Cases}
\zz{
The assurance case~\cite{Kelly1999Arguing}, also known as safety case, is an essential construct within safety-critical systems for demonstrating the safety and reliability of a system within specific operational contexts. These cases typically encompass aspects of system design, development, and maintenance, with an ultimate aim to ensure that the system meets safety and reliability criteria to achieve expected performance in real-world operation. The theoretical origin of assurance cases is traced to the domain of logical reasoning, notably introduced by the British philosopher Stephen Toulmin in 1958~\cite{toulmin1958uses}. The concept gained prominence with the rapid development in complex industries and the \yx{wide use} of novel automation technologies, as humans faced unprecedented technological risks~\cite{cleland2012evidence}. The evolution and widespread practical application of the assurance case were notably influenced by the 1988 Piper Alpha oil platform disaster~\cite{sklyar2020assurance}, underscoring the vital role of systematic, structured argumentation in assessing and establishing system safety in increasingly intricate and risk-prone technological landscapes.

Today, assurance cases, or safety cases, play a crucial role across various domains, particularly in industries that demand high standards of safety, reliability, and compliance. Representative application fields include:

\begin{itemize}
\item \textbf{Aerospace industry}~\cite{kelly1997building,rushby2015understanding}: Due to stringent safety requirements, aerospace engineering employs assurance cases to verify and assure the safety and reliability of airplanes~\cite{graydon2007assurance}, satellites~\cite{austin2017cubesat}, and spacecraft systems~\cite{vierhauser2019interlocking}.
\item \textbf{Railway industry}~\cite{medhurst2012safety,beugin2018safety}: \yx{A}ssurance cases are used to substantiate the safety and reliability of railway systems, such as signaling, train control, and operating equipment, reducing accident risk and ensuring passenger and staff safety.
\item \textbf{Automotive industry}~\cite{griessnig2017development,palin2010assurance}: With the advent of autonomous driving~\cite{bourbouh2021integrating}, assurance cases are deployed to argue \yx{for} the safety and reliability of self-driving systems.
\item \textbf{Medical devices}~\cite{bloomfield2012safety}: Medical device manufacturers (e.g., infusion pumps~\cite{larson2013open}, pacemakers~\cite{jee2010assurance}) utilize assurance cases to demonstrate the safety and compliance of the design, manufacturing, and usage processes of their products.
\item \textbf{Nuclear energy industry}~\cite{Bloomfield2009Safety,leveson2011use,wassyng2011software}: Given stringent demands for safety and compliance, assurance cases are employed to assess the safety of nuclear power stations, facilities, and nuclear material management systems.
\item \textbf{Oil and chemical industry}~\cite{henderson2012safety,baram2010preventing,mendes2014reforming}: In the oil, gas, and chemical sectors, assurance cases are utilized to evaluate and ensure safety and reliability throughout the process, preventing major accidents, averting environmental disasters, and safeguarding workers and environmental safety.
\item \textbf{Military and defense}~\cite{kelly2012safety}: In the highly security-sensitive military and defense sector, assurance cases are used to evaluate the safety and reliability of weapon systems, communication systems, and defensive mechanisms.
\item \textbf{Finance and banking}~\cite{duncan2014compliance}: Financial and banking industries leverage assurance cases to verify the security and compliance of financial transaction systems, safeguarding financial data and transactions.
\item \textbf{Safety management and regulation development}~\cite{bloomfield2017using}: In shaping safety management and regulations, such as cybersecurity regulation~\cite{bloomfield2017using}, school disaster prevention~\cite{widowati2021elementary}, and pandemic control policies~\cite{habli2020enhancing}, assurance cases play a role in risk assessment, design, and confirmation of control measures, provision of safety evidence, and promoting continuous improvement, thereby ensuring system safety and effective risk management.
\end{itemize}

The purpose of an assurance case is to articulate a clear, comprehensive, and dependable argument that a system's operation meets acceptable safety within a specific environment~\cite{Kelly1999Arguing}. An assurance case serves as a tool for communicating ideas and information, often conveying content to a third party such as regulatory authorities. To achieve this convincingly, it must be as \emph{clear} as possible. The \emph{system} referred to by an assurance case can be any object, such as a pipeline network, software configuration, or a set of operating procedures; the concept is not confined to considerations of traditional engineering ``design". Absolute safety is an unattainable goal, and the existence of an assurance case is to persuade others that the system is sufficiently safe, embodying \emph{acceptable safety} with tolerable risks. Safety argumentation must take into consideration premises, as nearly any system might be unsafe if used improperly or unexpectedly, such as arguing for the safety of conventional house bricks~\cite{kelly2004systematic}. Therefore, part of the work of an assurance case is defining the context or specific environment of safety. An assurance case consists of three main elements, namely goals, argumentation, and evidence, and the relationship between these three elements is depicted in Figure~\ref{ac_component}.

\begin{figure}[!t]
\centering
\begin{minipage}[c]{0.38\textwidth}
\centering
\includegraphics[width=0.65\textwidth]{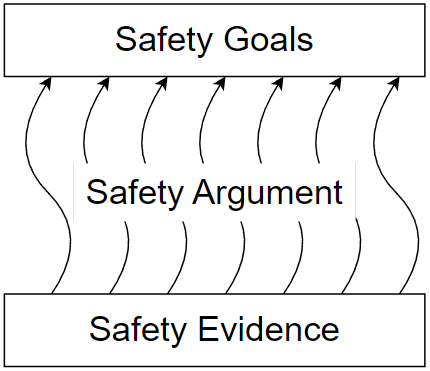}
\end{minipage}
\hspace{0.02\textwidth}
\begin{minipage}[c]{0.58\textwidth}
\centering
\includegraphics[width=0.8\textwidth]{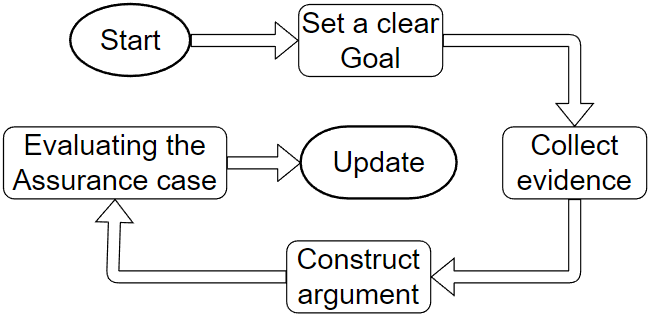}
\end{minipage}\\[3mm]
\begin{minipage}[t]{0.38\textwidth}
\centering
\caption{Structure of assurance cases.~\cite{Weaver2004The}}
\label{ac_component}
\end{minipage}
\hspace{0.02\textwidth}
\begin{minipage}[t]{0.58\textwidth}
\centering
\caption{Creation process of assurance cases.}
\label{ac_build}
\end{minipage}
\end{figure}

The process of creating an assurance case consists of four basic steps: identifying goals, gathering evidence, constructing arguments, and evaluating the assurance case~\cite{Bloomfield2020Assurance}. As shown in Figure~\ref{ac_build}, these steps build the fundamental framework of the assurance case, providing directions for safety engineers and project managers. This structured approach ensures a coherent and transparent connection between the goals, argumentation, and evidence, facilitating a clear and persuasive presentation of the system's safety and reliability. It is noteworthy that these four steps are not completed all at once but are iteratively performed throughout the project development process. As the project evolves and requirements change, the assurance case may need to be updated and modified. Furthermore, to ensure the quality and effectiveness of the assurance case, these four steps require good collaboration among the team members. This iterative and collaborative approach ensures that the assurance case remains aligned with the project's ongoing development and continues to reflect an accurate and robust representation of the system's safety and reliability.
}

\subsection{Large Language Models}

\zz{Large language models~\cite{cosler2023nl2spec} have their origins in the progressive evolution of machine learning algorithms and natural language processing techniques. They mark a significant advancement from traditional rule-based systems, employing deep learning architectures such as Transformers~\cite{vaswani2017attention}, introduced by Vaswani et al. in 2017. Application domains for these models are diverse, encompassing machine translation, text generation, sentiment analysis, summarization, and more. The implementation rationale of large language models lies in their ability to process and generate human-like text by learning from vast amounts of textual data, capturing intricate patterns and dependencies in language. Advantages of these models include their high versatility and adaptability across various tasks, often outperforming task-specific models. However, they are not without disadvantages; their large-scale nature demands extensive computational resources for both training and inference. Additionally, concerns regarding ethical considerations, biases embedded within the training data, and the potential lack of interpretability and transparency make the deployment and use of large language models a complex consideration.}

\zz{
Large language models are capable of accomplishing a wide range of tasks. Their utilization is straightforward, necessitating only an input box through which ``prompts" are sent to guide the model's responses. However, truly harnessing the full potential of these models is less straightforward. It requires a certain expertise in crafting these prompts. 

We have categorized several techniques for making effective use of large language models, as summarized in Table~\ref{table_llm}. These techniques include strategies to improve instruction quality, use of reference text, task decomposition, promoting the model's ``thinking'' process, integrating external tools, and systematic testing. For instance, Technique T1 (Being Specific) is a method to improve instruction quality by making queries more targeted, thereby eliciting more relevant responses from the model. Another example, Technique T12 (Solution Strategy) makes the model generate several potential solutions before coming up with a final answer, allowing it to explore various avenues of thought.
Furthermore, systematic testing plays an important role in the effective usage of language models. Techniques T17 and T18 involve comparing model outputs to gold standard answers and conducting A/B tests respectively, allowing for the evaluation and improvement of model performance. In short, these techniques collectively offer an approach to refine prompts and thereby extract more meaningful and valuable output from large language models.
Each technique listed in Table~\ref{table_llm} can be individually applied or combined with others, depending on the complexity of the task at hand and the specific objectives of the user.
}

\begin{table}[!t]
  \centering
  \caption{\zz{Classification and summary of usage techniques for large language models}}
  \begin{tabular}{|l|l|l|}
  \hline
  \textbf{Category} & \textbf{Technique} & \textbf{Technique ID} \\
  \hline
  Optimizing Instruction Quality & Being Specific & T1 \\
  \cline{2-3}
   & Role-play & T2 \\
  \cline{2-3}
   & Instruction Segmentation & T3 \\
  \cline{2-3}
   & Specifying Steps & T4 \\
  \cline{2-3}
   & Providing Examples & T5 \\
  \cline{2-3}
   & Setting Length & T6 \\
  \hline
  Leveraging Reference Text & Answer Reference & T7 \\
  \cline{2-3}
   & Citation Reference & T8 \\
  \hline
  Task Decomposition & Intent Classification & T9 \\
  \cline{2-3}
   & Information Filtering & T10 \\
  \cline{2-3}
   & Paragraph Summarization & T11 \\
  \hline
  Making the Model ``Think" & Solution Strategy & T12 \\
  \cline{2-3}
   & Simulate Thinking Process & T13 \\
  \cline{2-3}
   & Asking for Omissions & T14 \\
  \hline
  Combining External Tools & Embedding-based Search & T15 \\
  \cline{2-3}
   & Code Execution & T16 \\
  \hline
  Systematic Testing & Comparing to Gold Standard Answers & T17 \\
  \cline{2-3}
   & Conducting A/B Tests & T18 \\
  \hline
  \end{tabular}
  \label{table_llm}
\end{table}

\zz{
\begin{itemize}
    \item T1 (Being Specific): Make queries more targeted by providing the model with detailed information for more relevant answers.
    \item T2 (Role-play): Assign a role to the model within the query for more creative answers.
    \item T3 (Instruction Segmentation): Use delimiters to distinguish different parts in the query.
    \item T4 (Specifying Steps): List out the steps needed to complete the task to help the model generate accurate answers.
    \item T5 (Providing Examples): Assist the model in understanding requirements through examples.
    \item T6 (Setting Length): Specify the desired length of output in the query.
    \item T7 (Answer Reference): Allow the model to generate more accurate answers by referring to a specific text.
    \item T8 (Citation Reference): Instruct the model to quote specific parts from the reference text for more in-depth answers.
    \item T9 (Intent Classification): Decompose complex queries by analyzing the main objective in user queries.
    \item T10 (Information Filtering): For applications requiring long conversations, summarize or filter out previous dialogue, keeping only the key information.
    \item T11 (Paragraph Summarization): If dealing with long documents, split them into multiple paragraphs for summarization, and then combine these summaries.
    \item T12 (Solution Strategy): Make the model generate possible solutions before producing the final answer.
    \item T13 (Simulate Thinking Process): Allow the model to conduct an internal monologue, simulating a ``thinking" process.
    \item T14 (Asking for Omissions): Ask the model if it has omitted important information in the problem-solving process.
    \item T15 (Embedding-based Search): Use embedding-based search for effective knowledge retrieval.
    \item T16 (Code Execution): Leverage the model's code generation capability to perform calculations or call APIs.
    \item T17 (Comparing to Gold Standard Answers): Evaluate the quality of the model output by comparing it with preset gold standard answers.
    \item T18 (Conducting A/B Tests): Compare the effects of different prompts on the model output to find the most effective prompting strategy.
\end{itemize}
}

\zz{
In the process of generating TDT nodes using large language models, as discussed in Section~\ref{sec:arch}, the techniques outlined above have been utilized. 
}

\subsection{Constraint Solvers}

\zz{Constraint solvers~\cite{Jaffar1994Constraint} originated from the field of artificial intelligence and mathematical programming in the latter half of the 20th century, becoming an essential tool for solving problems expressed through constraints. The application fields of constraint solvers are manifold, including scheduling, planning, resource allocation, and various optimization problems. The implementation principle relies on techniques such as backtracking, consistency checking, and local search, often coupled with heuristics, to explore the solution space systematically and efficiently. Advantages of constraint solvers include their flexibility in modeling complex relationships and the ability to find optimal or near-optimal solutions. However, their disadvantages may involve high computational costs for large or complex problems and difficulty in modeling some real-world scenarios. For example, constraint solvers are widely used in airline scheduling, where constraints like the maximum number of working hours for pilots, mandatory rest periods, and aircraft maintenance schedules must be simultaneously satisfied. In this application, constraint solvers enable the creation of feasible schedules that adhere to all necessary regulations, though the complexity and size of the problem may present computational challenges.}

\section{Tool Architecture and Implementation}\label{sec:arch}
\begin{figure}[!t]
  \centering
  \begin{minipage}[c]{0.36\textwidth}
  \centering
  \includegraphics[width=1\textwidth]{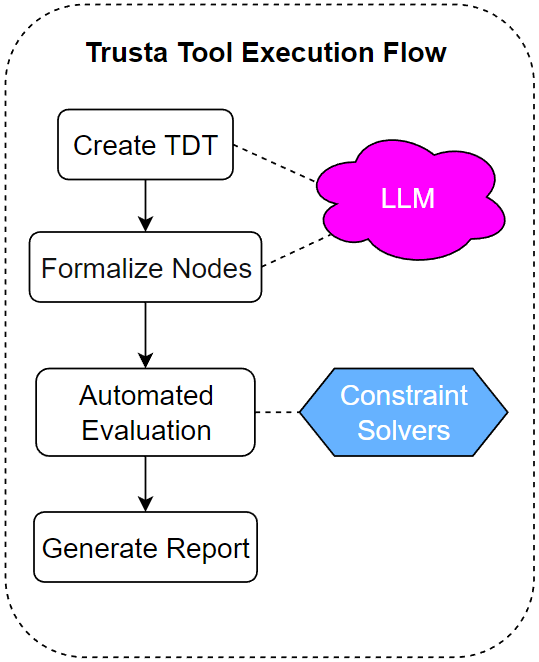}
  \end{minipage}
  \hspace{0.02\textwidth}
  \begin{minipage}[c]{0.6\textwidth}
  \centering
  \includegraphics[width=1\textwidth]{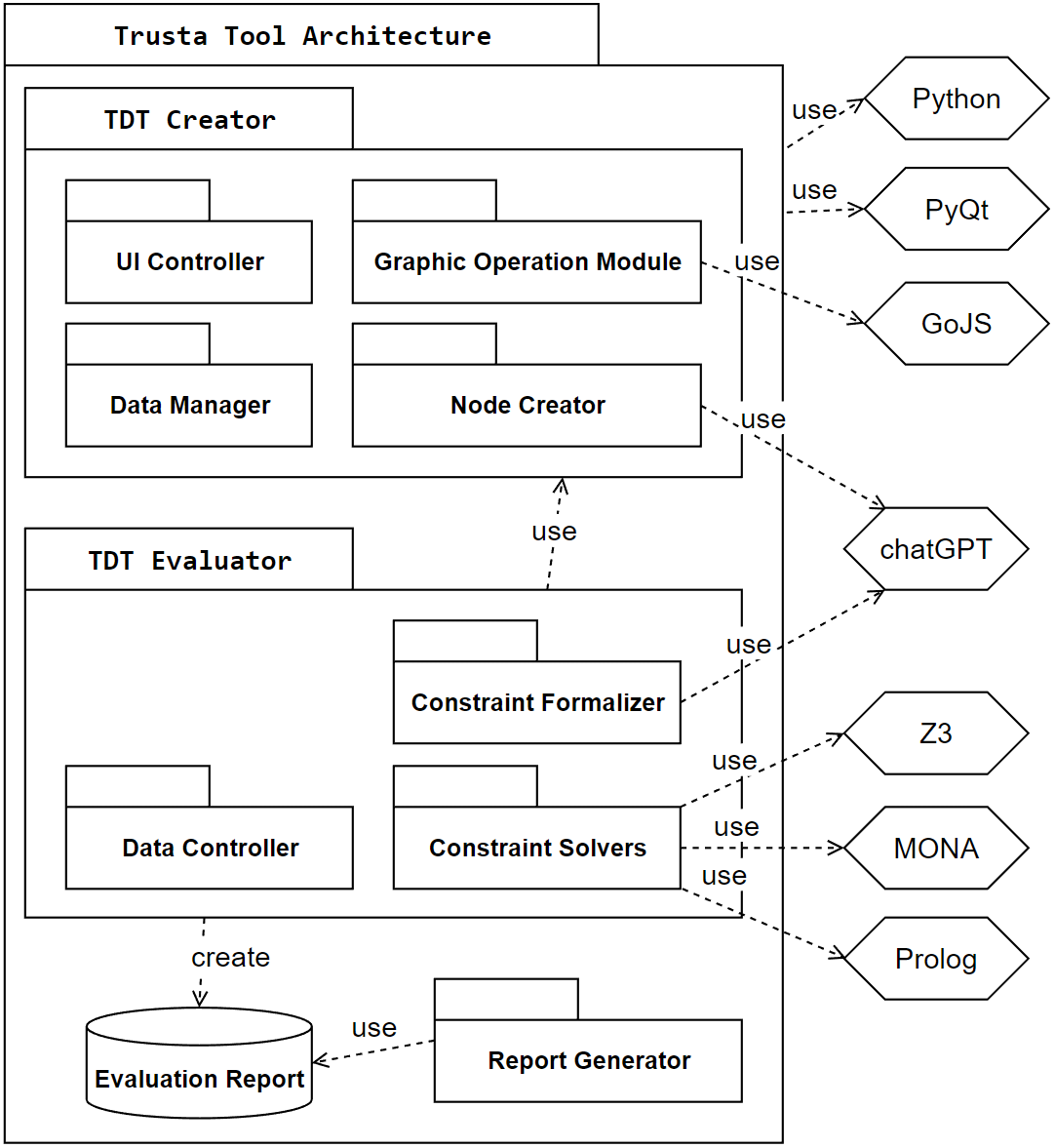}
  \end{minipage}
  \caption{\zz{An overview of the execution flow and the functional architecture of Trusta.}}
  \label{flow_arch}
  \end{figure}

In Figure~\ref{flow_arch}, we give an overview of the execution flow and the functional architecture of Trusta.
The tool is a desktop application created with Python's GUI library PyQt~\cite{Willman2021Overview}.
It can be used as an IDE to create TDTs, which are graphical representations of assurance cases, and provide various graphical transformation operations.
The tool consists of three modules: TDT Creator, TDT Evaluator, and Report Generator. Below we discuss each of them in more detail.

\subsection{TDT Creator}
The TDT Creator consists of four sub-modules: (1) a UI controller is in charge of responding to users' actions, (2) \zz{a node creates or utilizes a large language model to derive child nodes from the upper layer,} (3) a data manager can modify the data in a tree, (4) a graphic operation module uses the data of a tree to render TDT graphics and interactively modify the tree.

\begin{landscape}
\begin{figure}[!t]
\centering
\includegraphics[height=0.9\textheight]{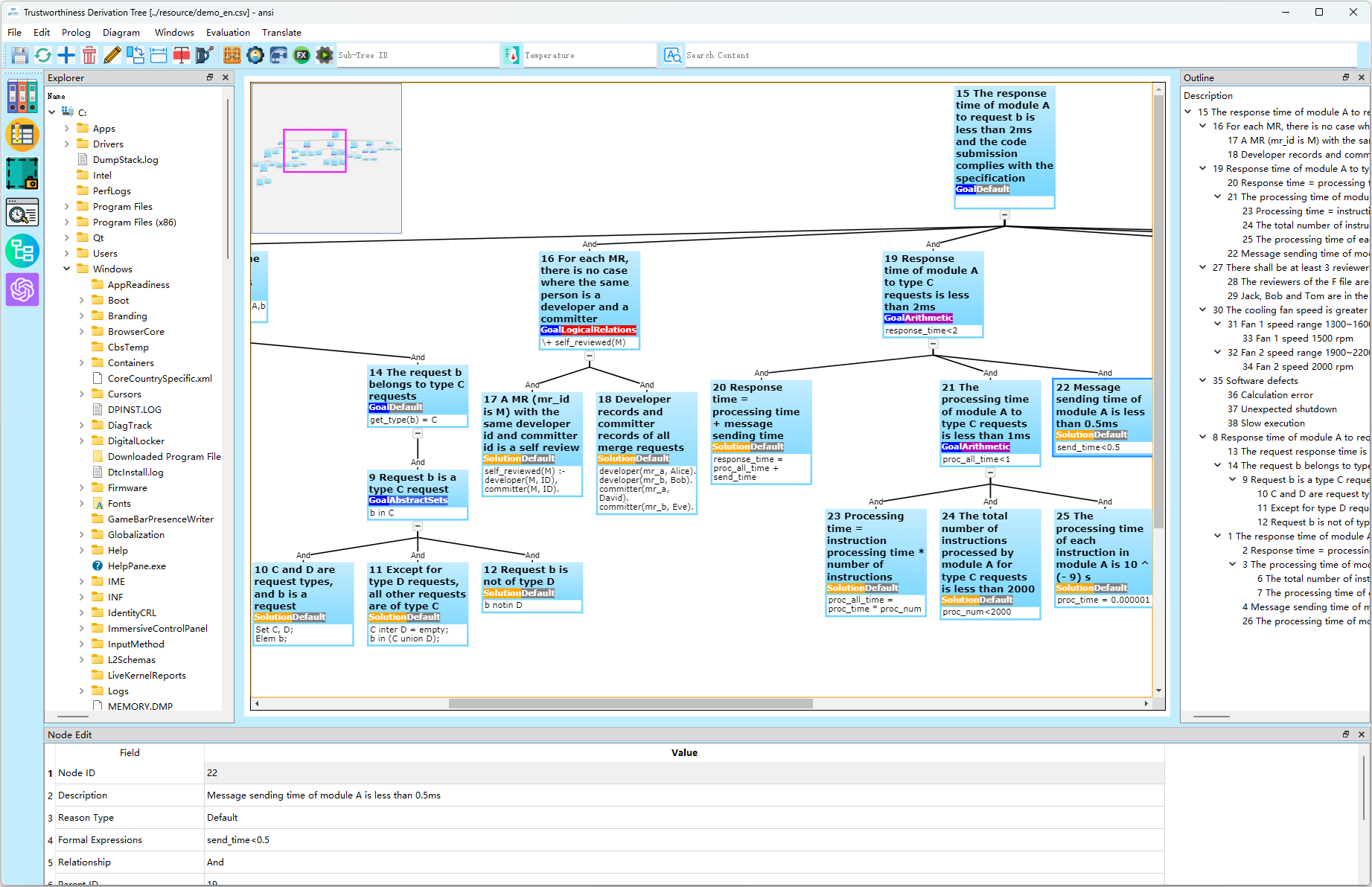}
\caption{A snapshot of Trusta}
\label{fig2}
\end{figure}
\end{landscape}

\paragraph{UI Controller} 
Figure~\ref{fig2} gives a snapshot of creating a TDT with Trusta.
After \zz{opening} a TDT, a tree is rendered automatically in the middle of the panel. Trusta provides many functions for editing and displaying the information of the nodes in the tree. For example, we can select, move, or resize nodes, modify node colors, rotate the entire tree, or hide some subtrees.
In the bottom of the panel, the information about a selected node is displayed and can be edited.
On the left of the panel is a project explorer, and on the right is
 an outline of the information with all the nodes in the TDT.

\paragraph{\zz{Node Creator}}
Trusta uses Prolog's syntax for Horn clauses that describe rules and axioms.
A rule can describe a two-level subtree, and multiple rules are able to describe a more complex multi-level tree.
Two examples are given in Figure~\ref{f:ruletext}.
On the left is a two-level tree generated by a rule, and on the right is a more complex tree generated by three rules. The module of text analyzer can recognize those rules so to construct TDTs on one hand and perform Prolog's inference on the other hand.
\zz{These rules governing the splitting of nodes can be formulated through manual procedures or with the integration of sophisticated large language models. Trusta, an innovative system in this context, seamlessly integrates such a large language model, thereby facilitating an efficient and user-friendly mechanism for node splitting. This integration enables users to accomplish node divisions with a single click. The resultant split is immediately usable and operational. Should any inconsistencies or errors be identified in the output, users are afforded the flexibility to enact manual adjustments. This control mechanism ensures that the TDT nodes generated align precisely with user expectations, thus providing a robust solution that harmonizes automated efficiency with user-guided precision. This blend of automated and manual control represents a significant advancement in the management of complex systems.}

\begin{figure}[!t]
\centering
\includegraphics[width=1\textwidth]{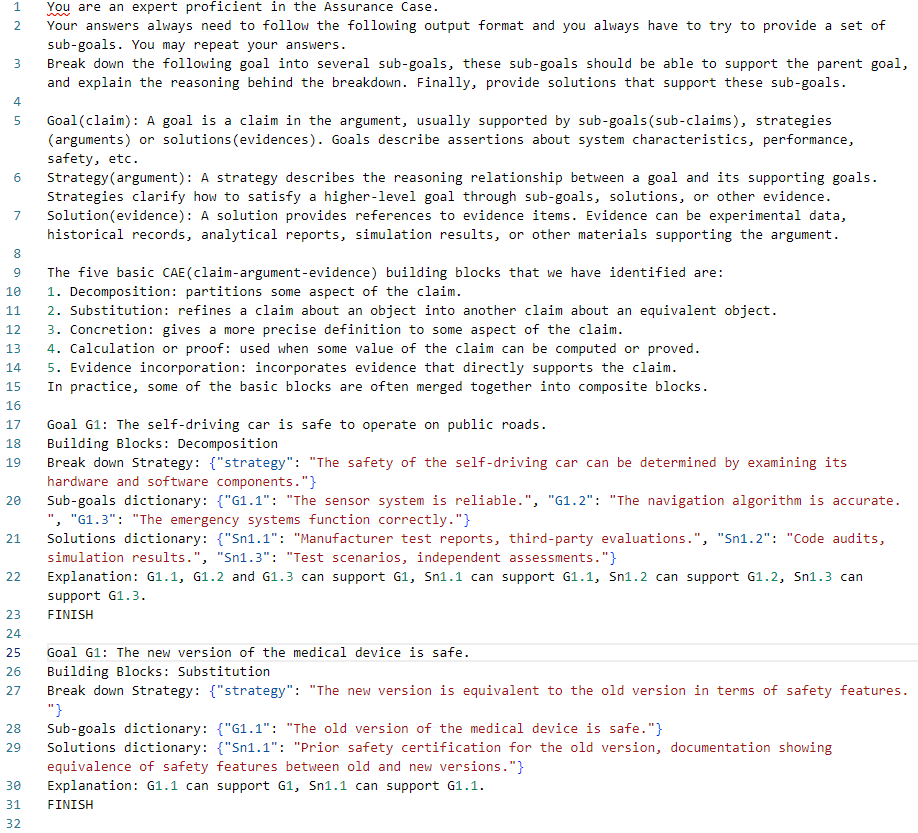}
\caption{Part 1 of 2: Prompt with domain knowledge of assurance case.}
\label{prompt_split_part1}
\end{figure}

\begin{figure}[!t]
\centering
\includegraphics[width=1\textwidth]{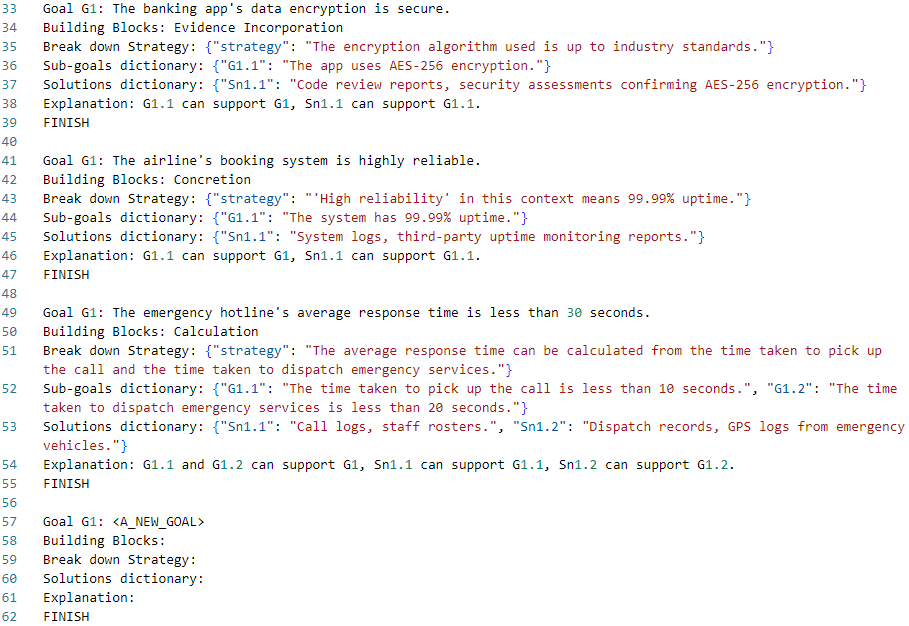}
\caption{Part 2 of 2: Prompt with examples of assurance case, including the final example awaiting completion by a large language model.}
\label{prompt_split_part2}
\end{figure}

\zz{The invocation of a large language model, particularly for complex tasks like assurance case generation, requires carefully crafted prompts. Due to the length of the prompts used for the creation of node-splitting rules, they are divided into two parts and illustrated in \yx{Figures}~\ref{prompt_split_part1} and \ref{prompt_split_part2}. The process under discussion is delineated into four distinct segments.}

\begin{figure}[!t]
\centering
\includegraphics[width=1\textwidth]{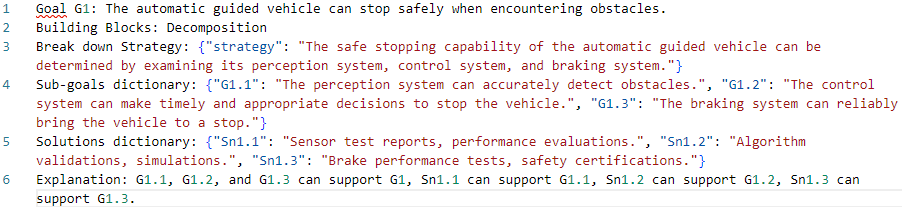}
\caption{\zz{Large language model output when splitting nodes.}}
\label{prompt_split_output}
\end{figure}

\zz{
\begin{enumerate}
\item The first segment (lines 1-3) sets the context and defines the role of the language model as an expert in assurance cases. It provides a general format that the model's output should follow and instructs the model to break down a given goal into various sub-goals. This section also asks the model to provide explanations for the breakdown as well as potential solutions for the sub-goals, setting up the stage for structured assurance case generation.
\item The second segment (lines 5-15) provides an in-depth look at the definitions and terminologies employed in assurance cases. This section not only defines what a ``Goal'', ``Strategy'', and ``Solution'' are but also outlines the five basic CAE (Claim-Argument-Evidence) building blocks~\cite{Netkachova2014Tool} essential for crafting assurance cases. These blocks are Decomposition, Substitution, Concretion, Calculation or Proof, and Evidence Incorporation. By introducing these conceptual tools, this segment equips the model with the necessary framework to understand and generate assurance cases more effectively.
\item The third segment (lines 17-55) offers multiple examples that individually highlight the use of each of the five building blocks: Decomposition, Substitution, Concretion, Calculation or Proof, and Evidence Incorporation. These examples cover various domains and goals such as self-driving cars, medical devices, and data encryption. For each example, the section details the building blocks employed, the breakdown strategy, the sub-goals, and solutions. Additionally, it provides explanations on how these elements are interconnected. These examples serve as both a comprehensive guide and a template for the model, aiding it in understanding how to structure and approach different types of assurance cases.
\item The fourth and final segment (lines 57-62) presents an incomplete example that consists solely of a placeholder for a goal, denoted as $\rm \langle A\_NEW\_GOAL \rangle$ , which is intended to be decomposed. This incomplete example follows the same format as the examples in the third segment and is designed for completion by a large language model. When invoking the model, $\rm \langle A\_NEW\_GOAL \rangle$  is replaced with a specific goal, as illustrated in the first line of Figure~\ref{prompt_split_output}.
\end{enumerate}
}

\zz{The model's output, as shown in Figure~\ref{prompt_split_output}, is then parsed by the Trusta tool to generate the TDT nodes. This effectively bridges the gap between theoretical modeling and practical implementation, demonstrating that the model's output is in a format compatible with Trusta for seamless integration into a workflow.}

\begin{figure}[!t]
\centering
\includegraphics[width=1\textwidth]{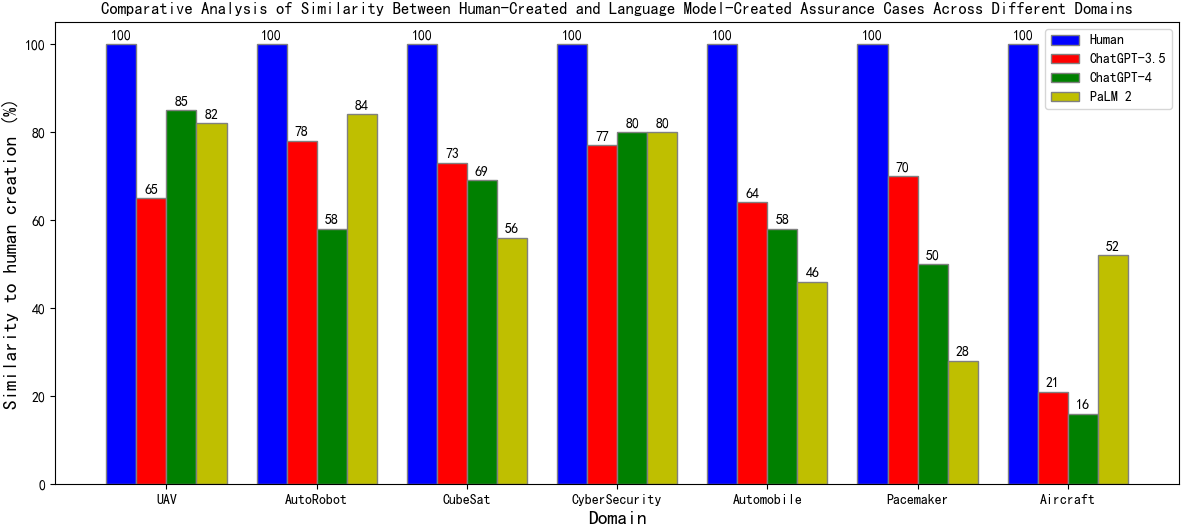}
\caption{\zz{Comparative analysis of similarity between human-created and language model-created assurance cases across different domains. The domains examined include UAV (Unmanned Aerial Vehicle)~\cite{vierhauser2019interlocking}, AutoRobot~\cite{bourbouh2021integrating}, CubeSat~\cite{austin2017cubesat}, CyberSecurity~\cite{bloomfield2017using}, Automobile~\cite{palin2010assurance}, Pacemaker~\cite{jee2010assurance}, and Aircraft~\cite{graydon2007assurance}. The models compared are ChatGPT-3.5, ChatGPT-4, and PaLM 2. Similarity is measured as a percentage of resemblance to human-created assurance cases in each domain.}}
\label{ComparativeSimilarity}
\end{figure}

\begin{figure}[t]
\begin{tabular}{cccccc}
	(a) &
$C :- C1, C2.$ &
$\raisebox{-.5\height}{\includegraphics[width=0.08\textwidth]{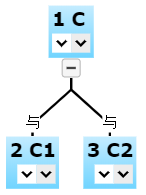}}$ &
\hspace{1cm}
(b) &
$\begin{bmatrix}
	C :- C1, C2. \\
	C1 :- C11, C12. \\
	C2 :- C21, C22, C23.
\end{bmatrix}$ &
$\raisebox{-.5\height}{\includegraphics[width=0.25\textwidth]{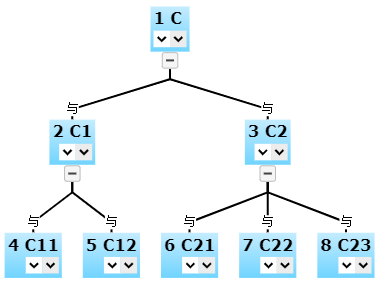}}$ 
\end{tabular}
\caption{Two examples of rule texts and TDT skeletons}
\label{f:ruletext}
\end{figure}

\zz{
In \yx{order} to evaluate the utility of cutting-edge language models for generating assurance cases, \yx{we conducted} a comprehensive analysis  on 57 assurance case fragments across seven distinct application domains. Figure~\ref{ComparativeSimilarity} presents the summarized results, comparing the semantic similarity between assurance cases created by humans and those generated by leading language models, namely ChatGPT-3.5 and ChatGPT-4 from OpenAI, as well as PaLM 2 from Google. The domains explored include Unmanned Aerial Vehicles (UAV)~\cite{vierhauser2019interlocking}, AutoRobot~\cite{bourbouh2021integrating}, CubeSat~\cite{austin2017cubesat}, CyberSecurity~\cite{bloomfield2017using}, Automobile~\cite{palin2010assurance}, Pacemaker~\cite{jee2010assurance}, and Aircraft~\cite{graydon2007assurance}. While the average similarity metrics generally lie between 50\%-80\%, this range still indicates a substantial contribution from these models in aiding the generation of assurance cases. It is worth mentioning that similarity here refers to the equivalence in the meaning of sentences within the assurance cases. Impressively, among the 57 fragments analyzed, 18 were found to have 100\% semantic similarity when generated by these AI models, illuminating their capability to produce \yx{reasonably accurate} assurance case content.}

\paragraph{Data Manager}
The Data Manager is mainly used to store and edit TDTs created from rule text \zz{or large language models}. It is involved when users add, delete, select, or modify TDT nodes. \zz{Typically}, a user \zz{begins by constructing} the skeleton of a TDT \zz{using a set of rules or the guidance from a large language model. Subsequently,} she \zz{refines} the content of each node by adding descriptions, types, and formal expressions. \zz{This results} in a complete TDT, \zz{capable of} represent\zz{ing} a normal assurance case\zz{, akin to the GSN or CAE notation}.

\paragraph{Graphic Operation Module}
The Graphic Operation Module is responsible for turning the TDT data stored by the Data Manager into diagrams and provides functions such as zooming, moving, and overview.
This module has contributed significantly to GoJS~\cite{Shahzad2016A}, a JavaScript library for creating interactive charts.
We embed browser controls on a PyQt based framework to run GoJS. 

\subsection{TDT Evaluator}
This is the module where formal methods are used for automatic reasoning about TDTs.
We use three constraint solvers~\cite{Rossi2008Constraint} to check the validity of the properties specified by the formal expression in each node of a TDT. Since different solvers are good at different types of reasoning, we use the \emph{Type} field in every node to indicate the evaluation type. For example, the type 'AbstractSet' in a node means that the formal expression in the node involves set operations about abstract sets, so we are going to employ MONA to solve the constraints. \zz{The process involves the translation of the natural language descriptions within nodes into formalized constraints, a task that can be undertaken through manual translation or through interactive translation with the assistance of a large language model~\cite{cosler2023nl2spec}.}

\begin{figure}[!t]
\centering
\includegraphics[width=1\textwidth]{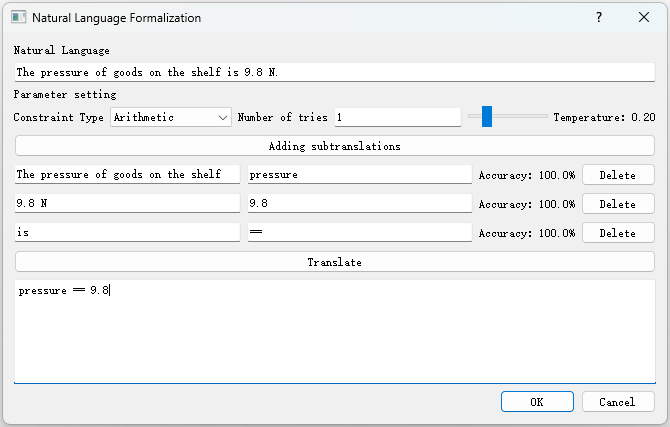}
\caption{\zz{Large language model interactive translation interface within Trusta.}}
\label{prompt_formal_UI}
\end{figure}

\paragraph{Data Controller}
In order to verify that the whole TDT is sound, it suffices to show the soundness of each two-level subtree in the TDT.
A two-level subtree consists of a parent node and several child nodes. It corresponds to a rule as shown in Figure~\ref{f:ruletext}. 
These child nodes represent the premises, and the parent node stands for the conclusion of the rule.
Suppose $F_1, F_2, ..., F_n$ are the formal expressions of premises, and $F$ is the formal expression of the conclusion. In addition, we allow two types of logical relations between the child nodes and their parent node, as indicated by a tag on each edge in Figure~\ref{demo_tdt}. The ``And" relation means that all the premises need to be combined to lead to the conclusion. In this case, we check if the formula $F_1 \wedge F_2 \wedge ... \wedge F_n \wedge \neg F$ is satisfiable. If it is unsatisfiable then the rule is sound. Otherwise, a solution exists and witnesses the unsoundness of the rule. The ``Or" relation means that any one of the premises can lead to the conclusion. In that case, we need to check the satisfiability of the formula $(F_1 \vee F_2 \vee ... \vee F_n) \wedge \neg F$.

\paragraph{Constraint \zz{Solvers}}
The satisfiability of the formulas given above are determined by constraint solvers. 
According to our experience with industrial case studies, we have summarized four types of constraints commonly encountered: logical relations, arithmetic, abstract sets, and concrete sets. 
Unfortunately, there exists no single solver that can solve all those types of constraints. Therefore, we have to call different solvers for different constraints. 
If the constraints are about logical relations, we resort to a lightweight Prolog built in Trusta. For arithmetic related to first-order theories, we take advantage of Z3. For some reasoning about abstract sets, i.e. unassigned sets whose elements are not explicitly known, we make use of MONA. For concrete sets whose elements are given in terms of arrays or lists, we use Python to deal with set operations.

\leaveout{ 
\begin{table}[t]
\caption{Constraint Solvers used in Trusta}
\begin{center}
\begin{tabular}{|c|c|}
\hline
\textbf{Types of constraints} & \textbf{Solvers} \\
\hline
Logical Relations & Prolog \\
\hline
Arithmetic & Z3 \\
\hline
Abstract Sets & MONA \\
\hline
Concrete Sets & Python \\
\hline
\end{tabular}
\label{tab2}
\end{center}
\end{table}
} 

Below we take a brief look at three types of constraints via a few simple examples. Consider the TDT shown in Figure~\ref{fig2}.
The number in the upper left corner of each node represents the node ID. The node IDs from the set $\{16,17,18\}$ correspond to a two-level subtree. The constraint for this subtree is captured by the expression $E_{Logical}$ in (\ref{eq1}). It is the conjunction of three parts: the first part says that a merge request with the same developer and committer is called self-reviewed; the second part is an evidence, a list of records showing the developers and committers of some merge requests; the third part is the negation of the property in the parent node, concerning about the absence of self-reviewed merge request, where the symbol `$\backslash +$' is the Prolog syntax for negation. The satisfiability of the formula $E_{Logical}$ can be checked by the lightweight Prolog built in Trusta.

\begin{equation}\label{eq1}\small\begin{array}{rcl}
	E_{Logical} & ~=~ & \ \
	``self\_reviewed(M) :-\ developer(M, ID), committer(M, ID)." \\
	& & \wedge \ 	``developer(mr_a, Alice). \\
	& & \qquad developer(mr_b, Bob). \\
	& & \qquad committer(mr_a, David). \\
	& & \qquad committer(mr_b, Eve)." \\
	& & \wedge \ \neg ``\backslash+ self\_reviewed(M)." 
\end{array}\end{equation}

Now consider the node IDs from the set $\{19,20,21,22\}$. They correspond to a two-level subtree whose constraints are about arithmetic and captured by the formula $E_{Arithmetic}$ in (\ref{eq2}).
The formula is a conjunction of four parts: the first part defines the relationship between the variables $response\_time$, $proc\_all\_time$, and $send\_time$; the second and third parts define the constraints on the last two variables; the last part is again the negation of the property in the parent node. The satisfiability of the formula $E_{Arithmetic}$ can be checked by  Z3.

\begin{equation}\label{eq2}
\small\begin{array}{rcl}
	E_{Arithmetic} & ~=~ & \ \
``response\_time=proc\_all\_time+send\_time" \\
& & \wedge \ ``proc\_all\_time<1" \\
& & \wedge \ ``send\_time<0.5" \\
& & \wedge \ \neg ``response\_time<2"
\end{array}\end{equation}

Then we consider the node IDs from the set $\{9,10,11,12\}$. They correspond to a two-level subtree that talks about abstract sets.
Their constraints are captured by the formula $E_{AbstractSet}$ in (\ref{eq3}). The formula is a conjunction of four parts: the first part defines the sets $C$ and $D$ together with an element $b$; the second and  third parts define the constraints between $C$, $D$, and $b$. The last part is the negation of the property in the parent node. We can employ MONA to check the satisfiability of the formula $E_{AbstractSet}$.

\begin{equation}\label{eq3}\small\begin{array}{rcl}
E_{AbstractSet} & ~=~ & \ \ 
``Set\ C, D; Elem\ b;" \\
& & \wedge \ ``C\ inter\ D = empty; b\ in\ (C\ union\ D);" \\
& & \wedge \ ``b\ notin\ D;" \\
& & \wedge \ \neg ``b\ in\ C;"
\end{array}\end{equation}

\begin{figure}[!t]
\centering
\includegraphics[width=1\textwidth]{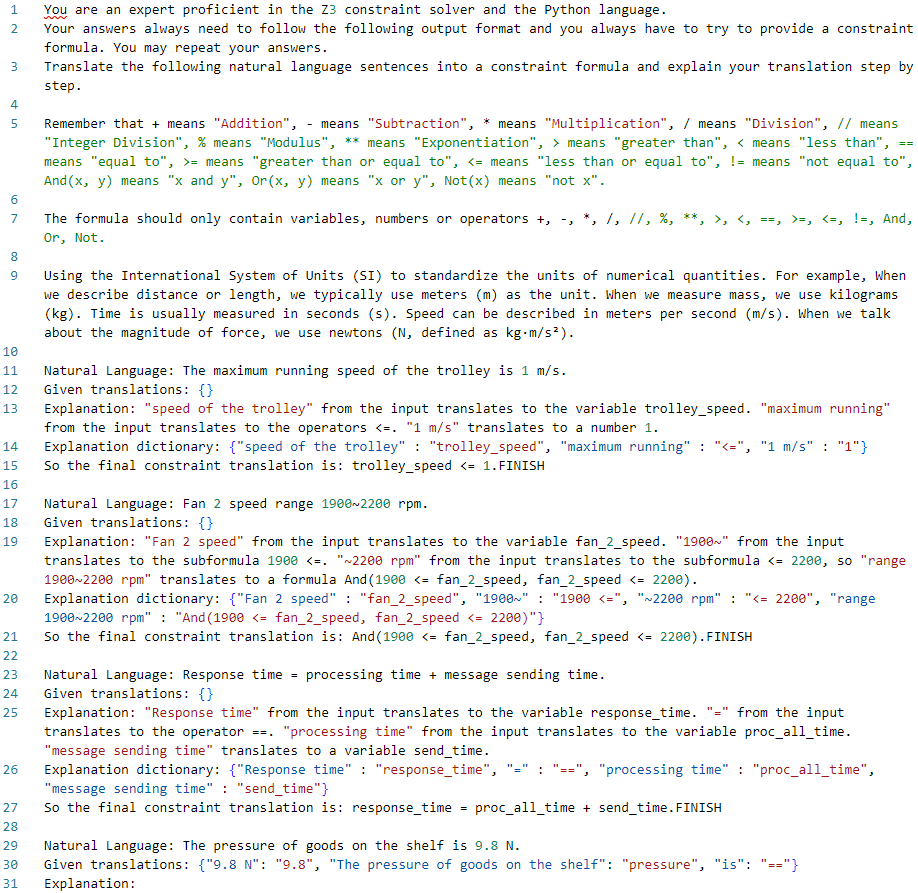}
\caption{\zz{Prompt of large language model translation from natural language to constraint expressions.}}
\label{prompt_formal_example}
\end{figure}

\begin{figure}[!t]
\centering
\includegraphics[width=1\textwidth]{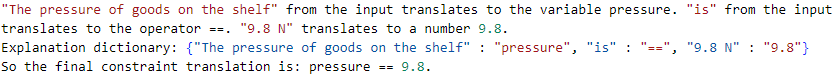}
\caption{\zz{Output of large language model translation from natural language to constraint expressions.}}
\label{prompt_formal_example_output}
\end{figure}

\paragraph{Constraint Formalizer}
\zz{The interactive translation interface within Trusta is illustrated in Figure~\ref{prompt_formal_UI}. The underlying conceptual framework draws inspiration from Cosler's work~\cite{cosler2023nl2spec} on translating natural language into temporal logics. We have made certain adaptations to the prompt words originally designed for translating temporal logics, as exemplified in Figure~\ref{prompt_formal_example}, in order to accommodate the transformation of natural language into constraint expressions.
We have revised the introduction of the problem context to focus on constraint expression considerations (lines 1-3). Symbol conventions have been adjusted to align with comprehensible notations for constraint solvers (lines 5-7). A novel provision regarding numeric units has been introduced, mandating a standardized adoption of international units (line 9). Furthermore, we present three illustrative examples of constraint translation challenges (lines 11-27). Conclusively, we furnish pending translations that encompass both natural language and manually generated sub-translation cues (lines 29-31). This framework is seamlessly extended by a large language model, adhering to the format of the provided examples, as demonstrated in Figure~\ref{prompt_formal_example_output}.
These adjustments facilitate the seamless transition from descriptive language to formal constraints, enhancing the applicability and efficacy of the translation process. This augmentation of the translation mechanism contributes to the broader goal of enhancing automated reasoning within the Trusta framework.}

\subsection{Report Generator}
Based on the results of constraint solving,  
Trusta reports on the vulnerabilities in the systems modeled by TDTs.
More specifically, if a property is invalid, the constraint solvers generate counterexamples to witness the invalidity of the property.
For example, if we change the third part of the formula $E_{Arithmetic}$ into $send\_time < 1.5$, then that formula is satisfiable. 
One solution is $(proc\_all\_time = 0.9, send\_time = 1.4, response\_time = 2.3)$.
In that case, the goal $response\_time < 2$ does not hold, so the TDT is unsound.
This kind of feedback from the constraint solvers provides TDT developers with more explicit information about the unsafe scenarios so they can quickly fix the problems.

\section{Case Studies}\label{sec:case}

Together with our industrial partner, we have constructed TDTs in more than a dozen real scenarios such as checking the consistency of software constructions and the trustworthiness of software implementation. Indeed, Trusta helped us to discover some subtle problems that were not noticed before.
\zz{In this section, we have conducted case demonstrations for both the creation and evaluation of TDT pertaining to automated guided vehicles (AGV) in warehouses.
This case uses the large language model ChatGPT-3.5~\cite{OpenAI2023GPT35}.
To assess the variances between different large language models and application domains in automatically generating assurance cases, we have also conducted experiments using three of the current leading models—ChatGPT-3.5, ChatGPT-4~\cite{OpenAI2023GPT4}, and PaLM 2~\cite{Google2023PaLM2}—across seven distinct domains for comparison.
}

\begin{figure}[!t]
\centering
\includegraphics[width=0.6\textwidth]{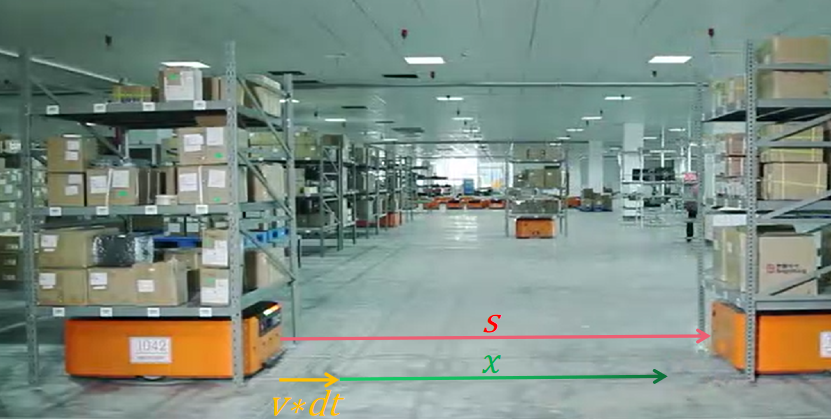}
\caption{AGV braking.}
\label{AGV_scenario}
\end{figure}

AGV robots move goods autonomously between the different areas of a warehouse, as shown in Figure~\ref{AGV_scenario}. They move along pre-designed routes and carry all kinds of loads. However, there are crossings between the route of one AGV and that of another AGV or the footway of a person. Therefore, potential risks exist and despite various preventive measures it is necessary to evaluate the trustworthiness of a warehouse with AGVs. We have constructed a TDT for this purpose.

\subsection{Creation of TDT}
\begin{figure}[!t]
\centering
\includegraphics[width=1\textwidth]{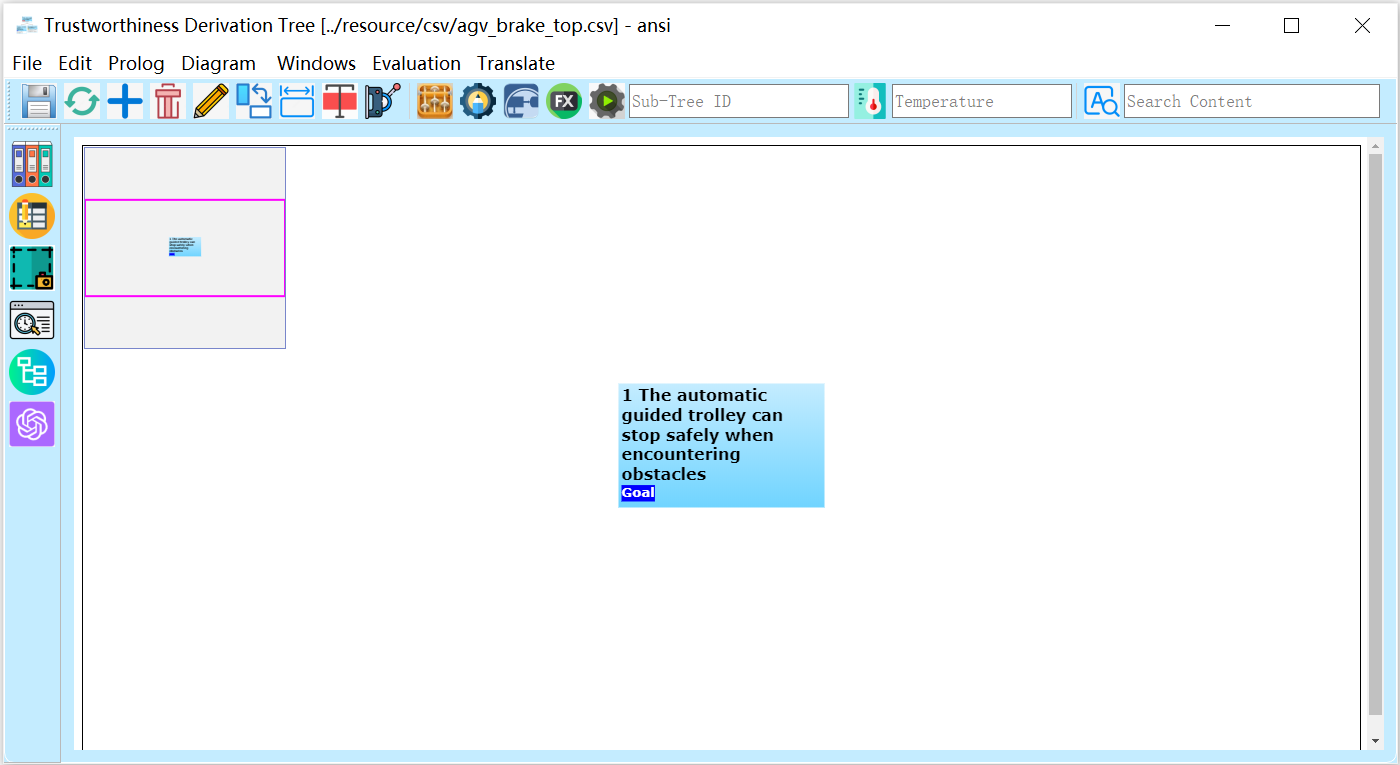}
\caption{\zz{Screenshot of the input for the creation of TDT using Trusta in the context of AGV.}}
\label{study_case_overview_input}
\end{figure}

\zz{
\yx{With} Trusta  for the construction of a TDT, it is only required to create a top-level goal, as illustrated in Figure~\ref{study_case_overview_input}. We established a goal node with the objective: ``The automatic guided vehicle can stop safely when encountering obstacles.'' Instructions were given to Trusta to decompose the goal into three layers, utilizing a language model's temperature parameter set at 0.8. This setting promotes greater creativity and enables the discovery of potentially overlooked subgoals. In the context of large language models~\cite{OpenAI2023Create}, the sampling temperature is a value ranging from 0 to 2. Higher values like 0.8 result in more random outputs, while lower values like 0.2 render the outputs more focused and deterministic.
}

\begin{figure}[!t]
\centering
\includegraphics[width=1\textwidth]{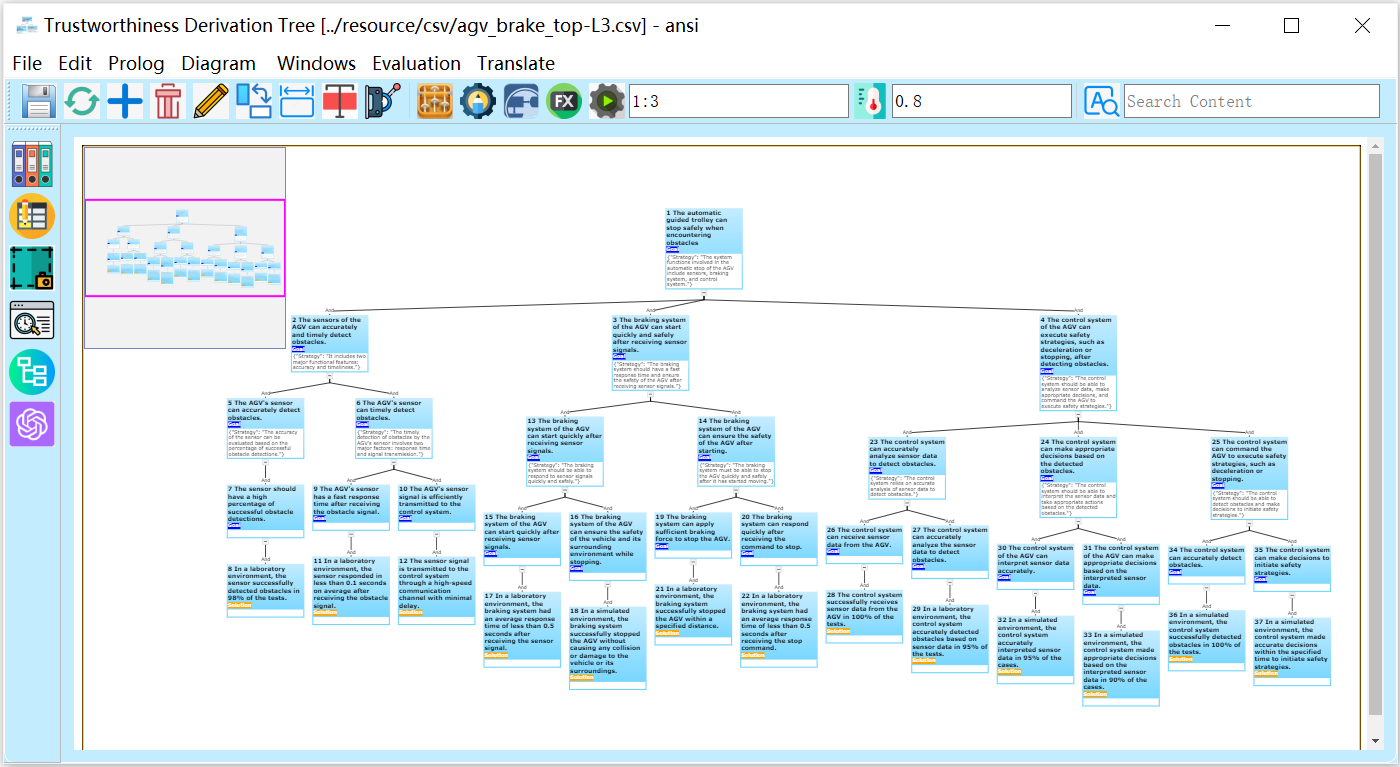}
\caption{\zz{Screenshot of the output for the creation of TDT using Trusta in the context of AGV.}}
\label{study_case_overview_output}
\end{figure}

\begin{figure}[!t]
\centering
\includegraphics[width=1\textwidth]{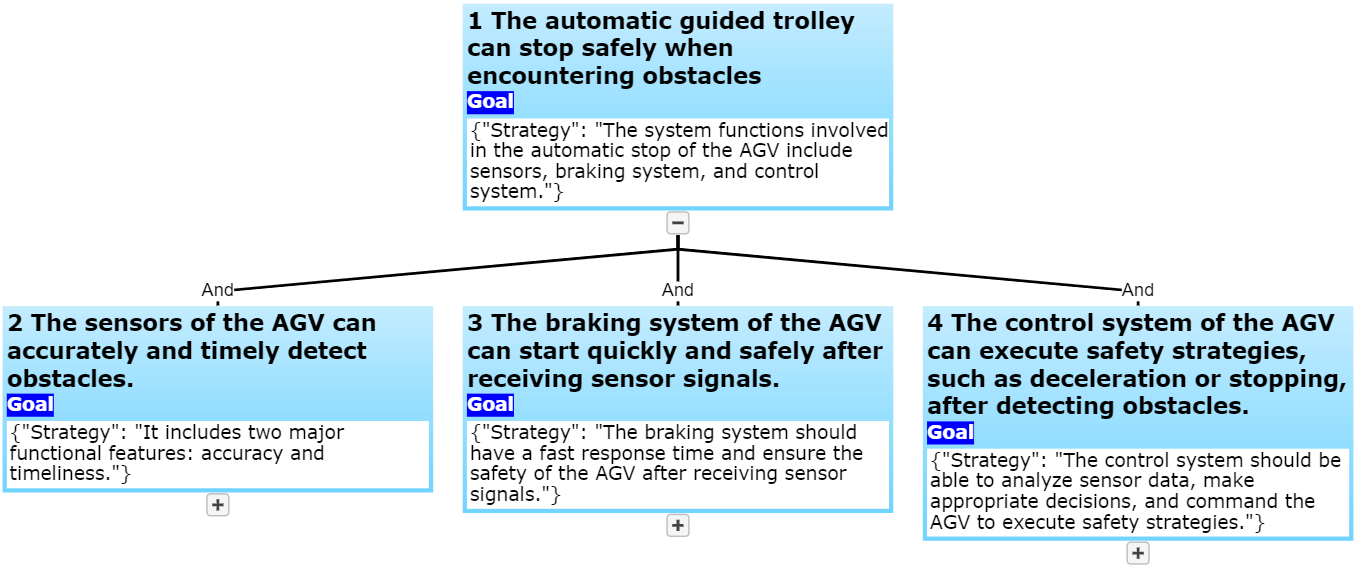}
\caption{\zz{Top-level node decomposition: The AGV can stop safely when encountering obstacles.}}
\label{study_case_top_1}
\end{figure}

\begin{figure}[!t]
\centering
\includegraphics[width=1\textwidth]{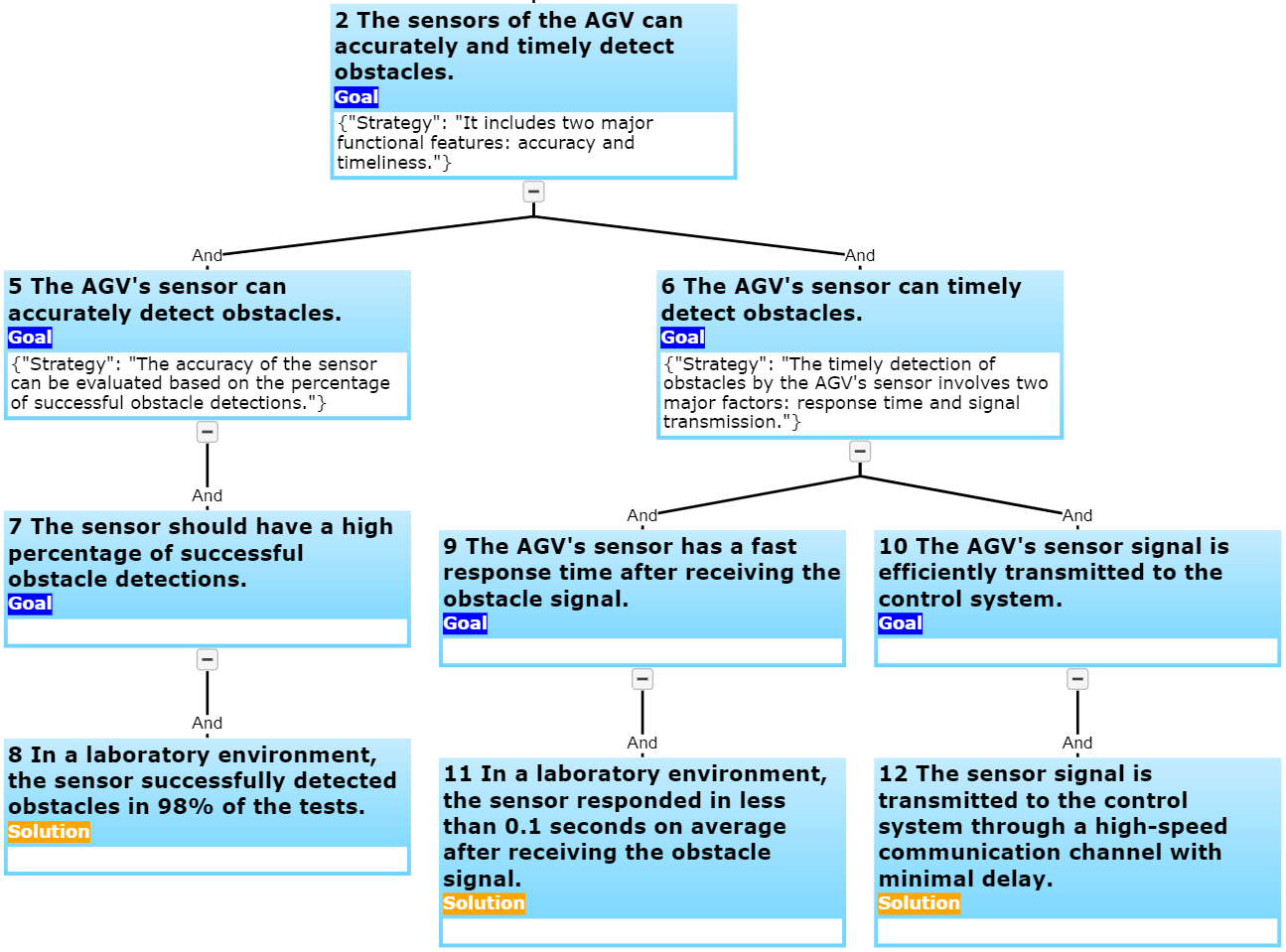}
\caption{\zz{Second-level node decomposition: The sensors of the AGV can accurately and timely detect obstacles.}}
\label{study_case_top_2}
\end{figure}

\begin{figure}
\centering
\includegraphics[width=\textwidth,height=7.1cm]{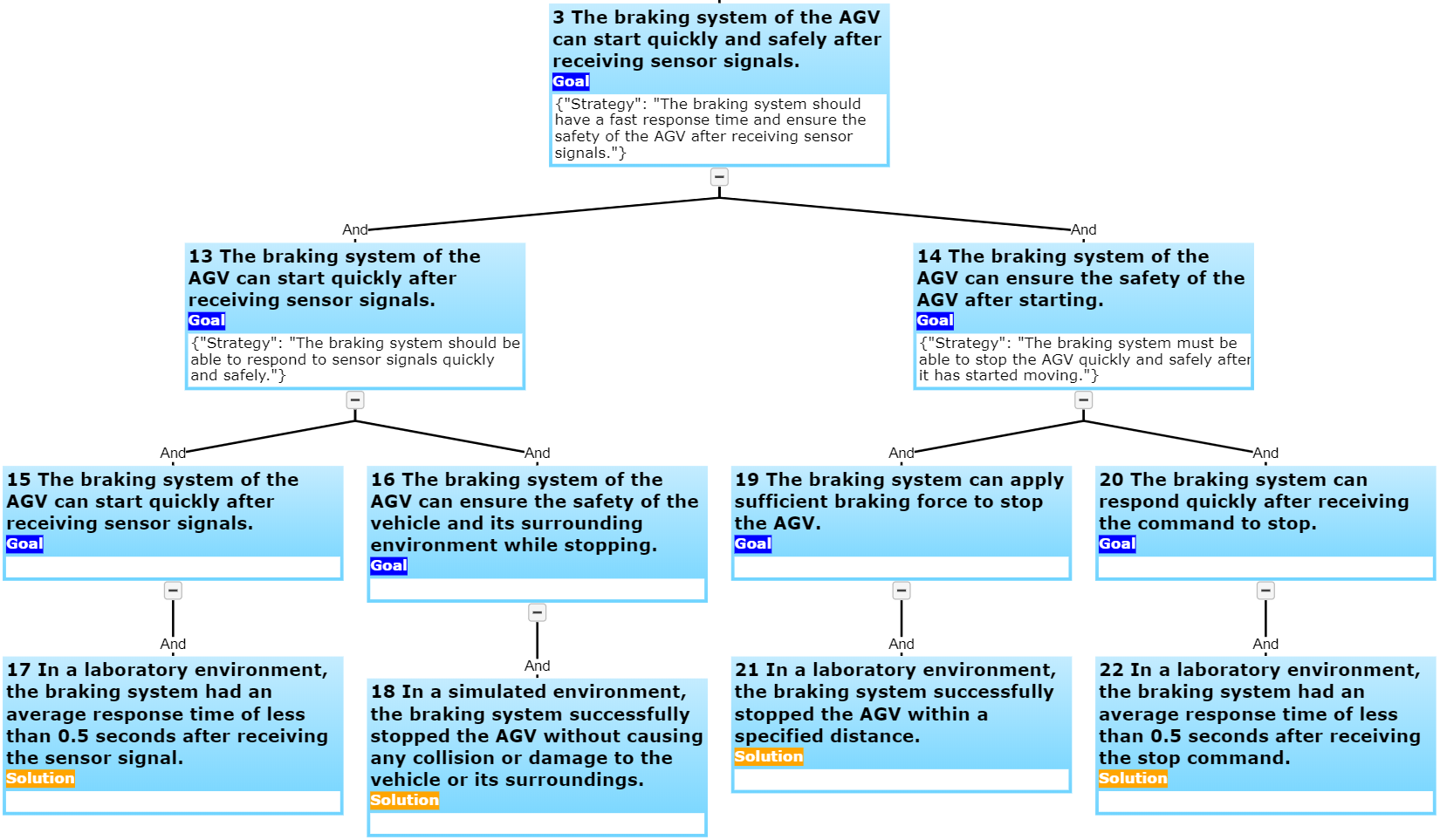}
\caption{\zz{Second-level node decomposition: The braking system of the AGV can start quickly and safely after receiving sensor signals.}}
\label{study_case_top_3}
\end{figure}

\begin{figure}
\centering
\includegraphics[width=1\textwidth,height=10.1cm]{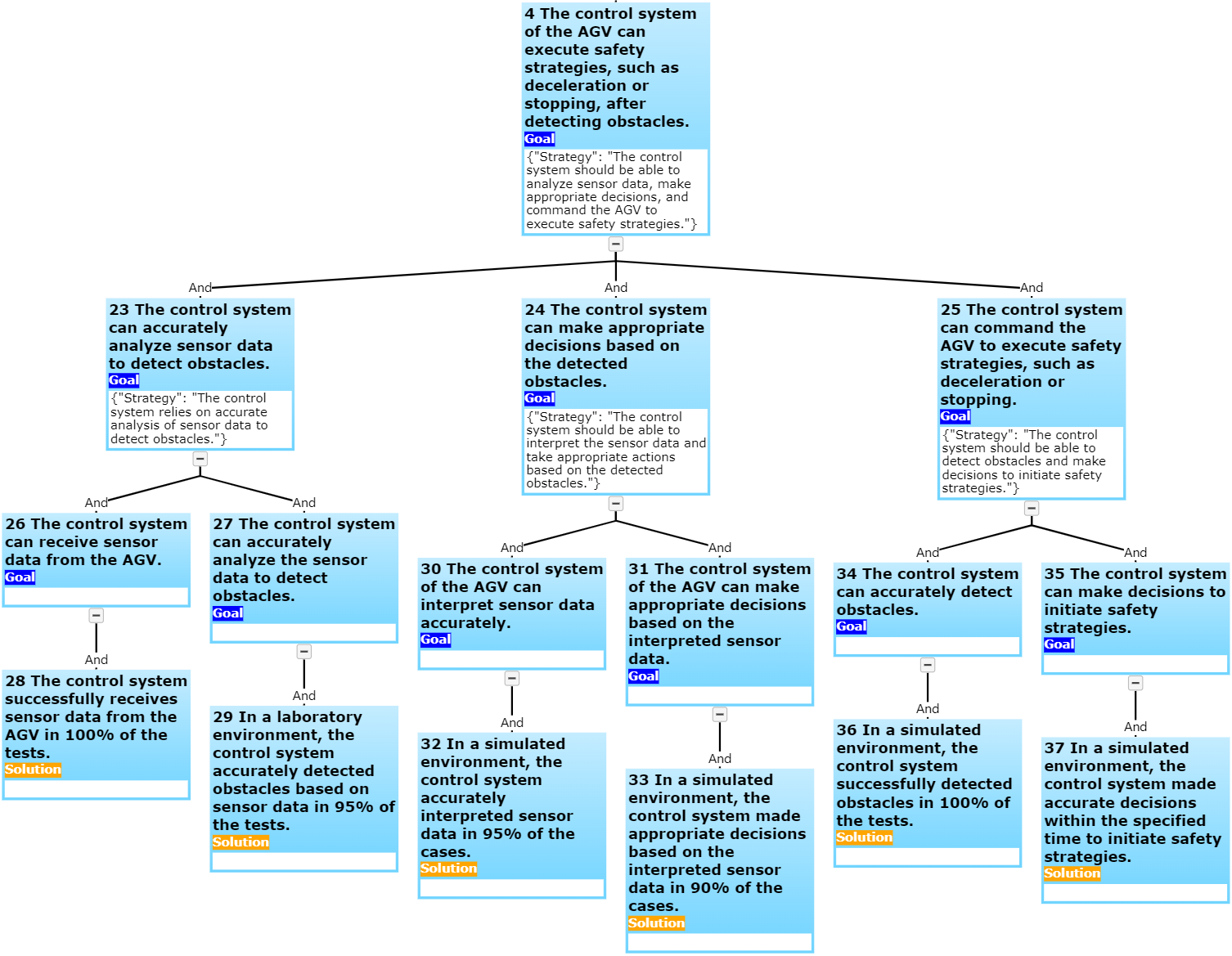}
\caption{\zz{Second-level node decomposition: The control system of the AGV can execute safety strategies, such as deceleration or stopping, after detecting obstacles.}}
\label{study_case_top_4}
\end{figure}

\zz{
Once the aforementioned inputs are prepared, Trusta is capable of generating a series of nodes, as depicted in Figure~\ref{study_case_overview_output}. Trusta automatically generated 36 nodes, encompassing 23 subgoal nodes and 13 solution nodes. Upon enlargement, these are respectively displayed in Figures~\ref{study_case_top_1}, \ref{study_case_top_2}, \ref{study_case_top_3}, and \ref{study_case_top_4}.
In Figure~\ref{study_case_top_1}'s decomposition, the top-level goal ``The automatic guided vehicle can stop safely when encountering obstacles'' (Node 1) has been broken down into three specific subgoals that form a comprehensive strategy to meet the main objective. The strategy defined for each subgoal elucidates the functionalities and considerations vital to the overarching aim of ensuring safe stopping of the AGV. These subgoals include the accurate and timely detection of obstacles by the AGV's sensors (Node 2), the rapid and safe initiation of the braking system after receiving sensor signals (Node 3), and the control system's capability to execute safety strategies, such as deceleration or stopping, after detecting obstacles (Node 4). Figures~\ref{study_case_top_2}, \ref{study_case_top_3} and \ref{study_case_top_4} follow the same pattern. Solutions at the leaf nodes of the TDT are created according to the upper-level nodes.

The above example illustrates the one-time multi-layer generation of TDT using Trusta. In practice, however, we can request the tool to decompose subgoals layer by layer, allowing users to make timely adjustments and further create more granular subgoals. As the decomposition progresses, there are typically two situations indicating that further decomposition of the goals might not be necessary: (1) when the generated nodes start to have meanings that are identical to their parent nodes or other existing nodes, and (2) when experts believe that the current goal node can be substantiated with evidence. This method of creating TDT aligns more closely with user expectations and ensures that the process does not consume excessive time.
}

\subsection{Evaluation of TDT}
Inevitably, AGVs traveling in a warehouse may encounter obstacles in front of them, which may be people, goods, or other AGVs.
A moving AGV should be able to recognize these obstacles and start to slow down and stop before collision.
In addition, the goods on the AGV should be stable without sliding.

Figure~\ref{AGV_scenario} shows the scenario in which an AGV is braking.
The AGV on the left is moving towards the right at speed $v$ and recognizes an obstacle with the distance of $s$ meters.
After $dt$ seconds of reaction time it starts to decelerate and brakes at a distance of $x$ meters.
In order to avoid a collision with the obstacle, the left AGV needs to generate sufficient deceleration.
However, if the deceleration is too large, it may cause the goods on the AGV to slide or even fall thus causing a safety hazard.

\begin{figure}[!t]
\centering
\includegraphics[width=1\textwidth,height=8.4cm]{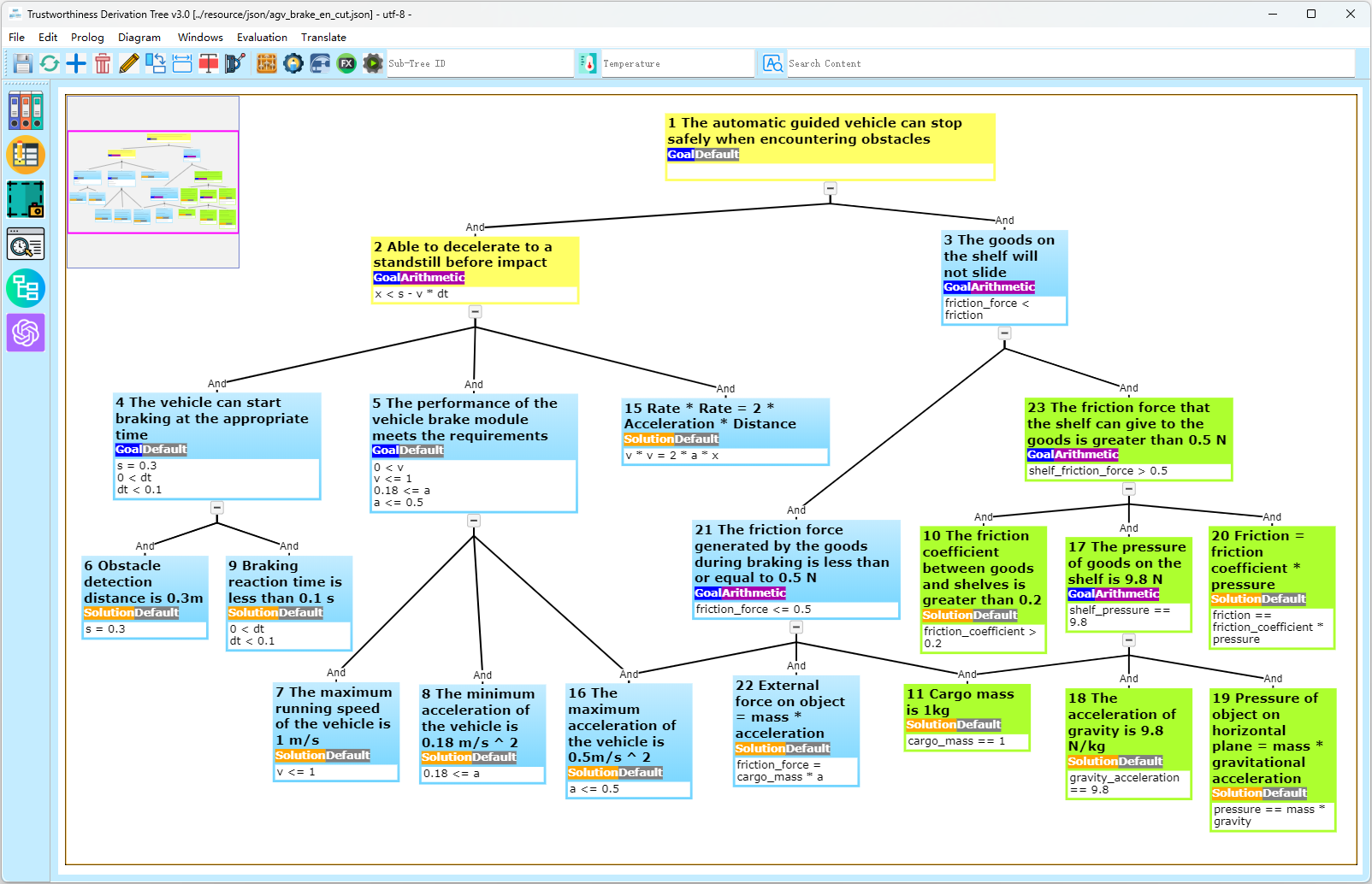}
\caption{\zz{Interactive adjustment of nodes in TDT creation: blue nodes represent ordinary nodes; green nodes symbolize newly generated constraint expression nodes; yellow nodes denote goals where subgoals cannot fully support the parent goal, indicating logical risks.}}
\label{study_case_UI}
\end{figure}

\begin{figure}[!t]
\centering
\includegraphics[width=1\textwidth,height=8.4cm]{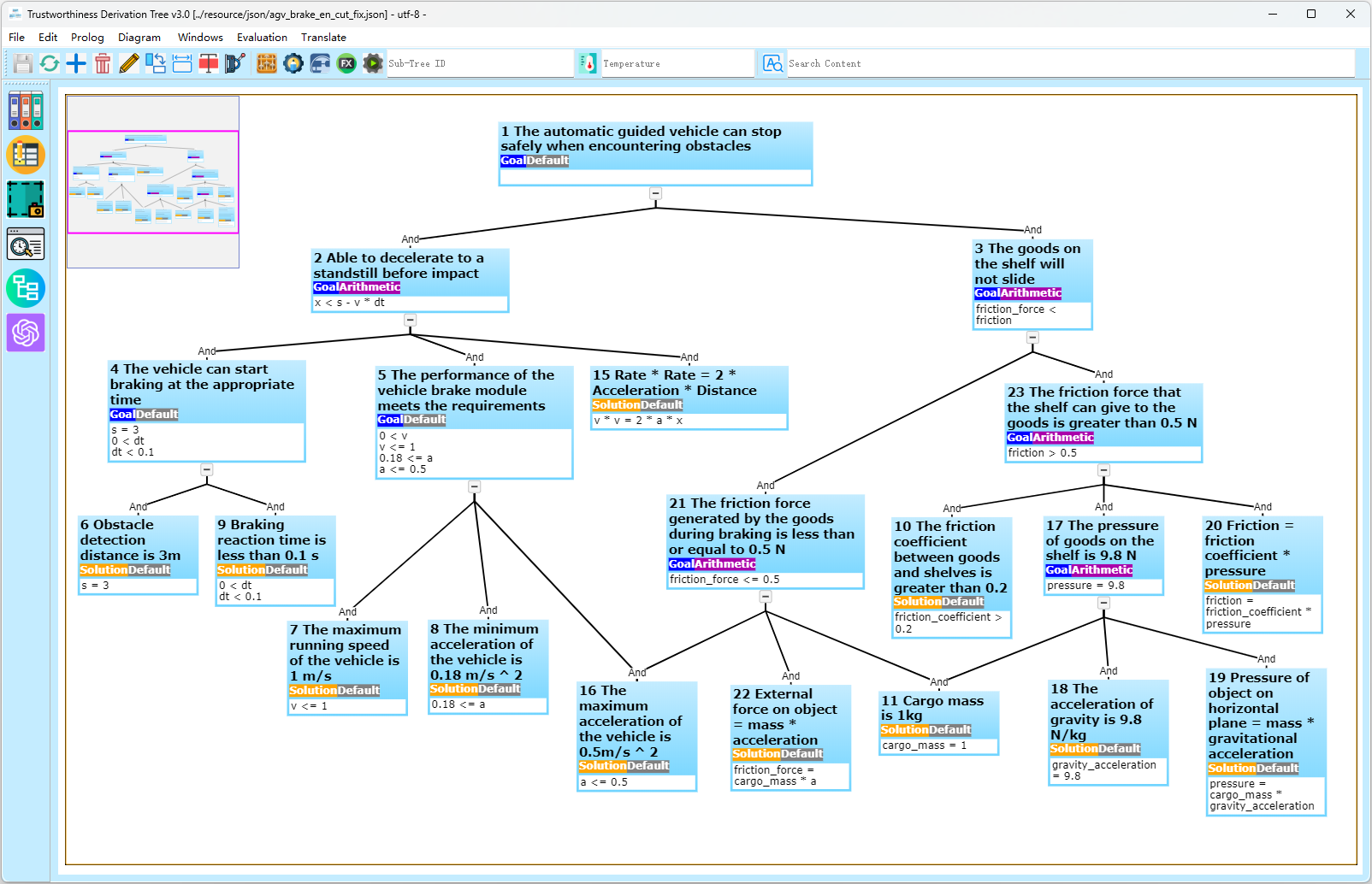}
\caption{\zz{Final TDT for the AGV's automatic braking case: Illustration of the refined structure after adjustments, emphasizing the translation result of constraint expressions in the evaluation stage.}}
\label{study_case_UI_after}
\end{figure}

\zz{
In collaboration with expert users, Trusta facilitated the creation of a TDT. The tool is capable of automatically translating clearly articulated node contents into constraint expressions and subsequently conducting \yx{formal reasoning with constraint solvers}. Nodes with ambiguous descriptions can be interactively adjusted by users, as depicted in Figure~\ref{study_case_UI}. Within the graphical representation, blue nodes denote ordinary nodes, green nodes represent newly generated constraint expression nodes, and yellow nodes are those identified by Trusta, with the assistance of a constraint solver, as logical risks — specifically, goals where subgoals do not entirely support the parent goal.

The case of AGV's automatic braking underwent adjustments, and the final resultant TDT is shown in Figure~\ref{study_case_UI_after}. It is noteworthy that the strategy information generated during the node creation phase is now concealed, shifting the focus during the evaluation stage more toward the translation process of constraint expressions. For a complete TDT \yx{with} auxiliary information, refer to Figure~\ref{AGV_GSN_ALL_TDT} in appendix~\ref{appendix:AGV_GSN}.
Table~\ref{nl2ce_result} provides a summary of the node translations depicted in Figure~\ref{study_case_UI_after}. These translations were accomplished through a large language model (GPT-3.5) converting natural language into constraint expressions. In Table 2, the ``Logical'' column has check marks indicating that the majority of these automated translations were logically coherent. However, the ``Variable'' column has check marks denoting that manual adjustments were often necessary for variable names to be compatible with the constraint solvers. It should be noted that the first five sentences in the dataset did not explicitly contain constraint information, causing the language model's translation efforts to fail. In these instances, manual creation was the only recourse to ensure correct solver execution.
}

\begin{landscape}
\begin{table}\small
\caption{\zz{Summary of node translations: comparing automated and manual approaches in constraint expression generation using GPT-3.5.}}
\begin{center}
\begin{tabular}{|c|p{6.5cm}|p{4.3cm}|c|c|p{4.3cm}|}
\hline
\textbf{No.} & \textbf{Natural language} & \textbf{LLM translation} & \textbf{Logical} & \textbf{Variable} & \textbf{Human adjustment results} \\
\hline
1 & The automatic guided trolley can stop safely when encountering obstacles & & & & \\
\hline
2 & Able to decelerate to a standstill before impact & & & & $x < s - v * dt$ \\
\hline
3 & The goods on the shelf will not slide & & & & $Fm < friction$ \\
\hline
4 & The trolley can start braking at the appropriate time & & & & $s = 3 ; 0 < dt ; dt < 0.1$ \\
\hline
5 & The performance of the trolley brake module meets the requirements & & & & $ 0 < v ; v <= 1 ; 0.18 <= a ; a <= 0.5 ; v * v = 2 * a * x $ \\
\hline
6 & Obstacle detection distance is 3m & $ obstacle\_distance == 3 $ & \Large \checkmark & & $ s = 3 $ \\
\hline
7 & The maximum running speed of the trolley is 1 m/s & $trolley\_speed <= 1$ & \Large \checkmark & & $ v <= 1 $ \\
\hline
8 & The minimum acceleration of the trolley is 0.18 $m/s^2$ & $trolley\_acceleration >= 0.18$ & \Large \checkmark & & $ 0.18 <= a $ \\
\hline
9 & Braking reaction time is less than 0.1 s & $braking\_reaction\_time < 0.1$ & & & $ 0 < dt ; dt < 0.1 $ \\
\hline
10 & The friction coefficient between goods and shelves is greater than 0.2 & $friction\_coefficient > 0.2$ & \Large \checkmark & \Large \checkmark & $ friction\_coefficient > 0.2 $ \\
\hline
11 & Cargo mass is 1kg & $cargo\_mass == 1$ & \Large \checkmark & \Large \checkmark & $ cargo\_mass = 1 $ \\
\hline
12 & Rate * Rate = 2 * Acceleration * Distance & $rate * rate == 2 * acceleration * distance$ & \Large \checkmark & & $ v * v = 2 * a * x $ \\
\hline
13 & The maximum acceleration of the trolley is $0.5m/s^2$ & $trolley\_acceleration <= 0.5$ & \Large \checkmark & & $ a <= 0.5 $ \\
\hline
14 & The pressure of goods on the shelf is 9.8 N & $shelf\_pressure == 9.8$ & \Large \checkmark & & $ pressure = 9.8 $ \\
\hline
15 & The acceleration of gravity is 9.8 N/kg & $gravity\_acceleration == 9.8$ & \Large \checkmark & \Large \checkmark & $ gravity\_acceleration = 9.8 $ \\
\hline
16 & Pressure of object on horizontal plane = mass * gravitational acceleration & $pressure == mass * gravity$ & \Large \checkmark & & $ pressure = cargo\_mass * gravity\_acceleration $ \\
\hline
17 & Friction = friction coefficient * pressure & $friction == friction\_coefficient * pressure$ & \Large \checkmark & \Large \checkmark & $ friction = friction\_coefficient * pressure $ \\
\hline
18 & The friction force generated by the goods during braking is less than 0.5 N & $friction\_force < 0.5$ & \Large \checkmark & & $ Fm <= 0.5 $ \\
\hline
19 & External force on object = mass * acceleration & $ext\_force == mass * accel$ & \Large \checkmark & & $ Fm = cargo\_mass * a $ \\
\hline
20 & The friction force that the shelf can give to the goods is greater than 0.5 N & $shelf\_friction\_force > 0.5$ & \Large \checkmark & & $ friction > 0.5 $ \\
\hline
\end{tabular}
\label{nl2ce_result}
\end{center}
\end{table}
\end{landscape} 



The fragment of a TDT shown in Figure~\ref{study_case_UI_after} aims to demonstrate that an AGV can safely brake when it encounters an obstacle. 
\zz{As previously mentioned, TDTs can be converted into assurance cases, and the GSN format corresponding to Figure~\ref{study_case_UI_after} can be specifically found in Figure~\ref{AGV_GSN_ALL} of appendix~\ref{appendix:AGV_GSN}.}
The top node has two subtrees: the left subtree argues that an AGV will not collide with obstacles, and the right one demonstrates that the goods on the AGV will not slide.
The left subtree is argued with the help of the uniformly variable linear motion equations, which are given in Table~\ref{tab3}.
The data for those parameters can be taken from the AGV's reference manual.
The maximum running speed is $v=1 m/s$, and the maximum deceleration is $a=0.5 m/s^2$.
The right subtree argues with the help of the equation for static friction.
Usually, the coefficient of friction between the goods and the shelf on top of the AGV is greater than $0.2$.


\begin{table}
\caption{Physical kinematic equations used in the AGV example}
\begin{center}
\begin{tabular}{|c|c|}
\hline
\textbf{Equation} & \textbf{Detail} \\
\hline
$s=vt$ & $s$ (displacement), $v$ (velocity), $t$ (time) \\
\hline
$v^2=2ax$ & $v$ (velocity), $a$ (acceleration), $x$ (displacement) \\
\hline
$F_N=mg$ & $F_N$ (normal force), $m$ (mass), $g$ (acceleration of gravity) \\
\hline
$F=uF_N$ & $F$ (sliding friction force), $u$ (frictional coefficient) \\
\hline
\end{tabular}
\label{tab3}
\end{center}
\end{table}

The development of the TDT uncovers some details that should be carefully considered. 
For example, on one hand we should set the minimum deceleration parameter of the AGV, as otherwise a collision may occur during braking, and on the other hand  it is more noteworthy to consider the materials of the goods' packaging and shelves. 
The coefficient of static friction of the corresponding material should exceed a certain value to ensure the stability of the goods.
Trusta turns out to be helpful for the tuning of parameters.
The construction and automatic evaluation of the TDT in this study case increase our confidence in the safe use of AGVs.

\section{Related Work}\label{sec:related}

Several assurance case editors have been developed to support GSN~\cite{Denney2012AdvoCATE,Matsuno2011D,2016ACEdit,2016CertWare,Voss2013Towards}. They facilitate the development and maintenance of assurance cases. Some of them offer assurance case patterns for users to reuse existing assurance cases~\cite{Matsuno2011D,Voss2013Towards}.
Luo et al.~\cite{Luo2017A} provided an excellent survey of assurance case tools and summarized a systematic process of assurance case assessment. They also developed a tool to facilitate human evaluation.
Chowdhury et al.~\cite{Chowdhury2020Systematic} proposed a set of rules that semi-formally define the structure and content of assurance cases.
These rules guide the work of assurance cases developers and reviewers.
Assurance cases developers are instructed to use a more rigorous approach to their arguments.
External reviewers have a basic checklist that guides them in assessing the rigor of arguments.
Maksimov et al.~\cite{Maksimov2019A} surveyed ten assurance case tools with evaluation capabilities.
These tools can examine both the structure and content of  assurance cases.
Structural checks include structural constraints, correctness, integrity checks, and user queries.
Content checks include argument evaluation, evidence evaluation, evaluation tracking, evaluation report, and evaluation interaction.
Different tools utilize different approaches for content checks such as type checking, Bayesian belief networks and Dempster-Shafer Theory. The only tool that uses a formal logic is Resolute~\cite{2016Resolute}. Similar to Trusta, Resolute is inspired by logic programming and accompanies claims with user-defined logical rules for formal analysis, but SMT solvers are not incorporated.

\zz{Among these tools, AdvoCATE~\cite{Denney2018Tool} stands out with a relatively higher degree of automation. It utilizes high-level argument patterns to assist in the assurance case creation process. By interpreting these templates, AdvoCATE can formulate detailed arguments, either interactively or through external data. Although its P-table structure effectively directs pattern instantiation, potential challenges may arise when dealing with intricate or non-conventional assurance scenarios, possibly affecting its versatility in diverse contexts.}

Although there are many types of assurance case tools, the current assurance case tools are still immature.
Most \zz{creation and} evaluation techniques they support still rely heavily on manual work.
The content and evidence in the assurance case are primarily in the form of natural language.
The validity of assurance case decomposition cannot be demonstrated.

In~\cite{Deng2021Trustworthiness}, we introduced TDTs as a more compact representation of assurance cases without losing their expressive power.
We gave a visualization tool that used Prolog syntax for importing and exporting TDT data. Basic soundness checking of TDTs cannot be carried out within the tool itself, but can be turned into the validity checking of propositional logical formulas and then performed by an external Prolog inference engine.

We note that the assessment of assurance cases plays a vital role in safety engineering.
Although some tools have been developed to assist assessors in judging the correctness of assurance cases, they are far from being sufficiently automated.
The accuracy of assessment is susceptible to human subjective factors.
\zz{The creation of assurance cases is largely a manual endeavor, further underscoring the low levels of automation in this domain.}
In addition, finding bugs and tweaking them after an assurance case is developed often waste a lot of time.

\section{Conclusion and Future Directions}\label{sec:con}
We have presented Trusta, a tool that allows for safety modeling and automatic validation, as well as a detailed report on safety vulnerabilities.
The TDTs created by this tool can be adapted from assurance cases by adding formal expressions, which can be used by constraint solvers to perform formal reasoning.
\zz{With the integration of large language models, Trusta also brings convenience in creating safety cases, and assists users in translating natural language into constraint expressions, streamlining the overall process.
In fact, within the Trusta tool, TDT and traditional GSN can be mutually converted. It can be observed that, without losing any information, the TDT representation is more compact, emphasizing key points, making it more easily readable.}
Our experiments with more than a dozen industrial cases show that Trusta is helpful to identify issues that are easily overlooked by manual inspection.

\zz{Looking forward to the future development of Trusta, several promising directions emerge.
First, there is an opportunity to trial and compare various large language models to discern the most effective ones for specific tasks among a few assurance cases. Such comparative studies may pave the way for nuanced insights and enhanced efficiencies.
Second, by integrating more theoretical knowledge, we can optimize prompt words to guide the models more effectively, harnessing their potential in a more targeted manner.
Third, the fine-tuning of these large language models to tailor their performance in specialized tasks is an exciting avenue for research. By customizing these models to the unique requirements of the safety domain, we anticipate significant advancements in their applicability and accuracy.
Finally, the integration and development of additional formal languages within Trusta will broaden the horizons of automatic reasoning within TDTs, making it more versatile and universally applicable.
These future endeavors signal a robust pathway towards more comprehensive, adaptable, and intelligent safety modeling and validation.}



\bibliographystyle{ACM-Reference-Format}
\bibliography{references}


\begin{thebibliography}{00}


\ifx \showCODEN    \undefined \def \showCODEN     #1{\unskip}     \fi
\ifx \showDOI      \undefined \def \showDOI       #1{#1}\fi
\ifx \showISBNx    \undefined \def \showISBNx     #1{\unskip}     \fi
\ifx \showISBNxiii \undefined \def \showISBNxiii  #1{\unskip}     \fi
\ifx \showISSN     \undefined \def \showISSN      #1{\unskip}     \fi
\ifx \showLCCN     \undefined \def \showLCCN      #1{\unskip}     \fi
\ifx \shownote     \undefined \def \shownote      #1{#1}          \fi
\ifx \showarticletitle \undefined \def \showarticletitle #1{#1}   \fi
\ifx \showURL      \undefined \def \showURL       {\relax}        \fi
\providecommand\bibfield[2]{#2}
\providecommand\bibinfo[2]{#2}
\providecommand\natexlab[1]{#1}
\providecommand\showeprint[2][]{arXiv:#2}

\bibitem[\protect\citeauthoryear{26262}{26262}{2011}]%
        {International2011Road}
\bibfield{author}{\bibinfo{person}{{ISO} 26262}.}
  \bibinfo{year}{2011}\natexlab{}.
\newblock \bibinfo{title}{Road Vehicles-Functional Safety}.
\newblock   (\bibinfo{year}{2011}).
\newblock
\showURL{%
\url{https://www.iso.org/standard/43464.html}}


\bibitem[\protect\citeauthoryear{ACEdit}{ACEdit}{2016}]%
        {2016ACEdit}
\bibfield{author}{\bibinfo{person}{ACEdit}.} \bibinfo{year}{2016}\natexlab{}.
\newblock   (\bibinfo{year}{2016}).
\newblock
\newblock
\shownote{https://code.google.com/p/acedit/.}


\bibitem[\protect\citeauthoryear{Austin, Mahadevan, Sierawski, Karsai,
  Witulski, and Evans}{Austin et~al\mbox{.}}{2017}]%
        {austin2017cubesat}
\bibfield{author}{\bibinfo{person}{Rebekah Austin},
  \bibinfo{person}{Nagabhushan Mahadevan}, \bibinfo{person}{Brian Sierawski},
  \bibinfo{person}{Gabor Karsai}, \bibinfo{person}{Arthur Witulski}, {and}
  \bibinfo{person}{John Evans}.} \bibinfo{year}{2017}\natexlab{}.
\newblock \showarticletitle{A CubeSat-payload radiation-reliability assurance
  case using goal structuring notation}. In \bibinfo{booktitle}{{\em In
  Proceedings of the 2017 Annual Reliability and Maintainability Symposium}}.
  \bibinfo{publisher}{IEEE}, \bibinfo{pages}{1--8}.
\newblock


\bibitem[\protect\citeauthoryear{Baram}{Baram}{2010}]%
        {baram2010preventing}
\bibfield{author}{\bibinfo{person}{Michael Baram}.}
  \bibinfo{year}{2010}\natexlab{}.
\newblock \bibinfo{booktitle}{{\em Preventing accidents in offshore oil and gas
  operations: the US approach and some contrasting features of the Norwegian
  approach}}.
\newblock \bibinfo{type}{{T}echnical {R}eport}. \bibinfo{institution}{Boston
  University School of Law}.
\newblock


\bibitem[\protect\citeauthoryear{Beugin, Legrand, Marais, Berbineau, and
  El-Koursi}{Beugin et~al\mbox{.}}{2018}]%
        {beugin2018safety}
\bibfield{author}{\bibinfo{person}{Julie Beugin}, \bibinfo{person}{Cyril
  Legrand}, \bibinfo{person}{Juliette Marais}, \bibinfo{person}{Marion
  Berbineau}, {and} \bibinfo{person}{El-Miloudi El-Koursi}.}
  \bibinfo{year}{2018}\natexlab{}.
\newblock \showarticletitle{Safety appraisal of GNSS-based localization systems
  used in train spacing control}.
\newblock \bibinfo{journal}{{\em IEEE Access\/}}  \bibinfo{volume}{6}
  (\bibinfo{year}{2018}), \bibinfo{pages}{9898--9916}.
\newblock


\bibitem[\protect\citeauthoryear{Bishop and Bloomfield}{Bishop and
  Bloomfield}{1998}]%
        {Bishop1998A}
\bibfield{author}{\bibinfo{person}{Peter Bishop} {and} \bibinfo{person}{Robin
  Bloomfield}.} \bibinfo{year}{1998}\natexlab{}.
\newblock \showarticletitle{A methodology for safety case development,
  Industrial Perspectives of Safety-Critical Systems}. In
  \bibinfo{booktitle}{{\em Proceedings of the sixth safety-critical systems
  symposium}}.
\newblock


\bibitem[\protect\citeauthoryear{Bloomfield and Bishop}{Bloomfield and
  Bishop}{2009}]%
        {Bloomfield2009Safety}
\bibfield{author}{\bibinfo{person}{Robin Bloomfield} {and}
  \bibinfo{person}{Peter Bishop}.} \bibinfo{year}{2009}\natexlab{}.
\newblock \showarticletitle{Safety and assurance cases: Past, present and
  possible future--an Adelard perspective}. In \bibinfo{booktitle}{{\em In
  Proceedings of the Making Systems Safer}}. \bibinfo{publisher}{Springer
  London}, \bibinfo{pages}{51--67}.
\newblock


\bibitem[\protect\citeauthoryear{Bloomfield, Bishop, Butler, and
  Netkachova}{Bloomfield et~al\mbox{.}}{2017}]%
        {bloomfield2017using}
\bibfield{author}{\bibinfo{person}{Robin Bloomfield}, \bibinfo{person}{Peter
  Bishop}, \bibinfo{person}{Eoin Butler}, {and} \bibinfo{person}{Kate
  Netkachova}.} \bibinfo{year}{2017}\natexlab{}.
\newblock \showarticletitle{Using an assurance case framework to develop
  security strategy and policies}. In \bibinfo{booktitle}{{\em In Proceedings
  of the Computer Safety, Reliability, and Security}}.
  \bibinfo{publisher}{Springer International Publishing},
  \bibinfo{pages}{27--38}.
\newblock


\bibitem[\protect\citeauthoryear{Bloomfield, Chozos, Cleland, and
  Adelard}{Bloomfield et~al\mbox{.}}{2012}]%
        {bloomfield2012safety}
\bibfield{author}{\bibinfo{person}{Robin Bloomfield}, \bibinfo{person}{Nick
  Chozos}, \bibinfo{person}{George Cleland}, {and} \bibinfo{person}{LLP
  Adelard}.} \bibinfo{year}{2012}\natexlab{}.
\newblock \showarticletitle{Safety case use within the medical devices
  industry}.
\newblock In \bibinfo{booktitle}{{\em Supplements to: Using safety cases in
  industry and healthcare}}. \bibinfo{publisher}{The Health Foundation},
  \bibinfo{address}{London}, \bibinfo{pages}{75--91}.
\newblock


\bibitem[\protect\citeauthoryear{Bloomfield and Rushby}{Bloomfield and
  Rushby}{2020}]%
        {Bloomfield2020Assurance}
\bibfield{author}{\bibinfo{person}{Robin Bloomfield} {and}
  \bibinfo{person}{John Rushby}.} \bibinfo{year}{2020}\natexlab{}.
\newblock \showarticletitle{Assurance 2.0: A manifesto}.
\newblock \bibinfo{journal}{{\em arXiv preprint arXiv:2004.10474\/}}
  (\bibinfo{year}{2020}).
\newblock


\bibitem[\protect\citeauthoryear{Bourbouh, Farrell, Mavridou, Sljivo, Brat,
  Dennis, and Fisher}{Bourbouh et~al\mbox{.}}{2021}]%
        {bourbouh2021integrating}
\bibfield{author}{\bibinfo{person}{Hamza Bourbouh}, \bibinfo{person}{Marie
  Farrell}, \bibinfo{person}{Anastasia Mavridou}, \bibinfo{person}{Irfan
  Sljivo}, \bibinfo{person}{Guillaume Brat}, \bibinfo{person}{Louise Dennis},
  {and} \bibinfo{person}{Michael Fisher}.} \bibinfo{year}{2021}\natexlab{}.
\newblock \showarticletitle{Integrating formal verification and assurance: an
  inspection rover case study}. In \bibinfo{booktitle}{{\em In Proceedings of
  the NASA Formal Methods}}. \bibinfo{publisher}{Springer International
  Publishing}, \bibinfo{pages}{53--71}.
\newblock


\bibitem[\protect\citeauthoryear{CertWare}{CertWare}{2016}]%
        {2016CertWare}
\bibfield{author}{\bibinfo{person}{CertWare}.} \bibinfo{year}{2016}\natexlab{}.
\newblock   (\bibinfo{year}{2016}).
\newblock
\newblock
\shownote{http://nasa.github.io/CertWare/.}


\bibitem[\protect\citeauthoryear{Chowdhury, Wassyng, Paige, and
  Lawford}{Chowdhury et~al\mbox{.}}{2020}]%
        {Chowdhury2020Systematic}
\bibfield{author}{\bibinfo{person}{Thomas Chowdhury}, \bibinfo{person}{Alan
  Wassyng}, \bibinfo{person}{Richard~F Paige}, {and} \bibinfo{person}{Mark
  Lawford}.} \bibinfo{year}{2020}\natexlab{}.
\newblock \showarticletitle{Systematic evaluation of (safety) assurance cases}.
  In \bibinfo{booktitle}{{\em International Conference on Computer Safety,
  Reliability, and Security}}. \bibinfo{publisher}{Springer},
  \bibinfo{pages}{18--33}.
\newblock


\bibitem[\protect\citeauthoryear{Cleland, Sujan, Habli, and Medhurst}{Cleland
  et~al\mbox{.}}{2012}]%
        {cleland2012evidence}
\bibfield{author}{\bibinfo{person}{George Cleland},
  \bibinfo{person}{Mark-Alexander Sujan}, \bibinfo{person}{Ibrahim Habli},
  {and} \bibinfo{person}{John Medhurst}.} \bibinfo{year}{2012}\natexlab{}.
\newblock \bibinfo{booktitle}{{\em Evidence: using safety cases in industry and
  healthcare}}.
\newblock \bibinfo{publisher}{The Health Foundation}. 1--32 pages.
\newblock


\bibitem[\protect\citeauthoryear{Cosler, Hahn, Mendoza, Schmitt, and
  Trippel}{Cosler et~al\mbox{.}}{2023}]%
        {cosler2023nl2spec}
\bibfield{author}{\bibinfo{person}{Matthias Cosler},
  \bibinfo{person}{Christopher Hahn}, \bibinfo{person}{Daniel Mendoza},
  \bibinfo{person}{Frederik Schmitt}, {and} \bibinfo{person}{Caroline
  Trippel}.} \bibinfo{year}{2023}\natexlab{}.
\newblock \showarticletitle{nl2spec: Interactively Translating Unstructured
  Natural Language to Temporal Logics with Large Language Models}. In
  \bibinfo{booktitle}{{\em International Conference on Computer Aided
  Verification}}. \bibinfo{publisher}{Springer}.
\newblock


\bibitem[\protect\citeauthoryear{Council}{Council}{2007}]%
        {National2007Software}
\bibfield{author}{\bibinfo{person}{National~Research Council}.}
  \bibinfo{year}{2007}\natexlab{}.
\newblock \bibinfo{booktitle}{{\em Software for dependable systems: Sufficient
  evidence?}}
\newblock \bibinfo{publisher}{National Academies Press}.
\newblock


\bibitem[\protect\citeauthoryear{De~Moura and Bj{\o}rner}{De~Moura and
  Bj{\o}rner}{2008}]%
        {De2008Z3}
\bibfield{author}{\bibinfo{person}{Leonardo De~Moura} {and}
  \bibinfo{person}{Nikolaj Bj{\o}rner}.} \bibinfo{year}{2008}\natexlab{}.
\newblock \showarticletitle{Z3: An efficient {SMT} solver}. In
  \bibinfo{booktitle}{{\em International conference on Tools and Algorithms for
  the Construction and Analysis of Systems}}. \bibinfo{publisher}{Springer},
  \bibinfo{pages}{337--340}.
\newblock


\bibitem[\protect\citeauthoryear{Deng, Chen, Du, Mao, Liang, Lin, and Li}{Deng
  et~al\mbox{.}}{2021}]%
        {Deng2021Trustworthiness}
\bibfield{author}{\bibinfo{person}{Yuxin Deng}, \bibinfo{person}{Zezhong Chen},
  \bibinfo{person}{Wenjie Du}, \bibinfo{person}{Bifei Mao},
  \bibinfo{person}{Zhizhang Liang}, \bibinfo{person}{Qiushi Lin}, {and}
  \bibinfo{person}{Jinghui Li}.} \bibinfo{year}{2021}\natexlab{}.
\newblock \showarticletitle{Trustworthiness Derivation Tree: A Model of
  Evidence-Based Software Trustworthiness}. In \bibinfo{booktitle}{{\em
  Proceedings of the 21st International Conference on Software Quality,
  Reliability and Security Companion (QRS-C)}}. \bibinfo{publisher}{IEEE},
  \bibinfo{pages}{487--493}.
\newblock


\bibitem[\protect\citeauthoryear{Denney and Pai}{Denney and Pai}{2018}]%
        {Denney2018Tool}
\bibfield{author}{\bibinfo{person}{Ewen Denney} {and} \bibinfo{person}{Ganesh
  Pai}.} \bibinfo{year}{2018}\natexlab{}.
\newblock \showarticletitle{Tool support for assurance case development}.
\newblock \bibinfo{journal}{{\em Automated Software Engineering\/}}
  \bibinfo{volume}{25}, \bibinfo{number}{3} (\bibinfo{year}{2018}),
  \bibinfo{pages}{435--499}.
\newblock


\bibitem[\protect\citeauthoryear{Denney, Pai, and Pohl}{Denney
  et~al\mbox{.}}{2012}]%
        {Denney2012AdvoCATE}
\bibfield{author}{\bibinfo{person}{Ewen Denney}, \bibinfo{person}{Ganesh Pai},
  {and} \bibinfo{person}{Josef Pohl}.} \bibinfo{year}{2012}\natexlab{}.
\newblock \showarticletitle{AdvoCATE: An assurance case automation toolset}. In
  \bibinfo{booktitle}{{\em International Conference on Computer Safety,
  Reliability, and Security}}. \bibinfo{publisher}{Springer},
  \bibinfo{pages}{8--21}.
\newblock


\bibitem[\protect\citeauthoryear{DO-178C}{DO-178C}{2011}]%
        {RTCA2011Software}
\bibfield{author}{\bibinfo{person}{DO-178C}.} \bibinfo{year}{2011}\natexlab{}.
\newblock \bibinfo{title}{Software Considerations in Airborne Systems and
  Equipment Certification}.
\newblock   (\bibinfo{year}{2011}).
\newblock
\showURL{%
\url{https://www.do178.org/}}


\bibitem[\protect\citeauthoryear{Duncan and Whittington}{Duncan and
  Whittington}{2014}]%
        {duncan2014compliance}
\bibfield{author}{\bibinfo{person}{Bob Duncan} {and} \bibinfo{person}{Mark
  Whittington}.} \bibinfo{year}{2014}\natexlab{}.
\newblock \showarticletitle{Compliance with standards, assurance and audit:
  does this equal security?}. In \bibinfo{booktitle}{{\em In Proceedings of the
  7th International Conference on Security of Information and Networks}}.
  \bibinfo{publisher}{Association for Computing Machinery},
  \bibinfo{pages}{77--84}.
\newblock


\bibitem[\protect\citeauthoryear{Google}{Google}{2023}]%
        {Google2023PaLM2}
\bibfield{author}{\bibinfo{person}{Google}.} \bibinfo{year}{2023}\natexlab{}.
\newblock \bibinfo{title}{Introducing {PaLM 2}}.
\newblock   (\bibinfo{year}{2023}).
\newblock
\newblock
\shownote{https://ai.google/discover/palm2/.}


\bibitem[\protect\citeauthoryear{Graydon, Knight, and Strunk}{Graydon
  et~al\mbox{.}}{2007}]%
        {graydon2007assurance}
\bibfield{author}{\bibinfo{person}{Patrick Graydon}, \bibinfo{person}{John
  Knight}, {and} \bibinfo{person}{Elisabeth Strunk}.}
  \bibinfo{year}{2007}\natexlab{}.
\newblock \showarticletitle{Assurance based development of critical systems}.
  In \bibinfo{booktitle}{{\em In Proceedings of the 37th Annual IEEE/IFIP
  International Conference on Dependable Systems and Networks}}.
  \bibinfo{publisher}{IEEE}, \bibinfo{pages}{347--357}.
\newblock


\bibitem[\protect\citeauthoryear{Griessnig and Schnellbach}{Griessnig and
  Schnellbach}{2017}]%
        {griessnig2017development}
\bibfield{author}{\bibinfo{person}{Gerhard Griessnig} {and}
  \bibinfo{person}{Adam Schnellbach}.} \bibinfo{year}{2017}\natexlab{}.
\newblock \showarticletitle{Development of the 2nd Edition of the ISO 26262}.
  In \bibinfo{booktitle}{{\em In Proceedings of the Systems, Software and
  Services Process Improvement}}. \bibinfo{publisher}{Springer International
  Publishing}, \bibinfo{pages}{535--546}.
\newblock


\bibitem[\protect\citeauthoryear{Group}{Group}{2021}]%
        {2021The}
\bibfield{author}{\bibinfo{person}{The Assurance Case~Working Group}.}
  \bibinfo{year}{2021}\natexlab{}.
\newblock \bibinfo{title}{Goal Structuring Notation Community Standard Version
  3}.
\newblock   (\bibinfo{year}{2021}).
\newblock
\newblock
\shownote{https://scsc.uk/SCSC-141C.}


\bibitem[\protect\citeauthoryear{Habli, Alexander, Hawkins, Sujan, McDermid,
  Picardi, and Lawton}{Habli et~al\mbox{.}}{2020}]%
        {habli2020enhancing}
\bibfield{author}{\bibinfo{person}{Ibrahim Habli}, \bibinfo{person}{Rob
  Alexander}, \bibinfo{person}{Richard Hawkins}, \bibinfo{person}{Mark Sujan},
  \bibinfo{person}{John McDermid}, \bibinfo{person}{Chiara Picardi}, {and}
  \bibinfo{person}{Tom Lawton}.} \bibinfo{year}{2020}\natexlab{}.
\newblock \showarticletitle{Enhancing Covid-19 Decision-Making by Creating an
  Assurance Case for Simulation Models}.
\newblock \bibinfo{journal}{{\em arXiv preprint arXiv:2005.08381\/}}
  (\bibinfo{year}{2020}).
\newblock


\bibitem[\protect\citeauthoryear{Henderson}{Henderson}{2012}]%
        {henderson2012safety}
\bibfield{author}{\bibinfo{person}{Jamie Henderson}.}
  \bibinfo{year}{2012}\natexlab{}.
\newblock \showarticletitle{Safety case use in the petrochemical industry}.
\newblock In \bibinfo{booktitle}{{\em Supplements to: Using safety cases in
  industry and healthcare}}. \bibinfo{publisher}{The Health Foundation},
  \bibinfo{address}{London}, \bibinfo{pages}{55--64}.
\newblock


\bibitem[\protect\citeauthoryear{{ISO/IEC 15026}}{{ISO/IEC 15026}}{2011}]%
        {International2011Systems}
\bibfield{author}{\bibinfo{person}{{ISO/IEC 15026}}.}
  \bibinfo{year}{2011}\natexlab{}.
\newblock \bibinfo{title}{Systems and Software Engineering-Systems and Software
  Assurance-{P}art 2: Assurance Case}.
\newblock   (\bibinfo{year}{2011}).
\newblock
\showURL{%
\url{https://www.iso.org/standard/52926.html}}


\bibitem[\protect\citeauthoryear{Jaffar and Maher}{Jaffar and Maher}{1994}]%
        {Jaffar1994Constraint}
\bibfield{author}{\bibinfo{person}{Joxan Jaffar} {and}
  \bibinfo{person}{Michael~J Maher}.} \bibinfo{year}{1994}\natexlab{}.
\newblock \showarticletitle{Constraint logic programming: A survey}.
\newblock \bibinfo{journal}{{\em The journal of logic programming\/}}
  \bibinfo{volume}{19} (\bibinfo{year}{1994}), \bibinfo{pages}{503--581}.
\newblock


\bibitem[\protect\citeauthoryear{Jee, Lee, and Sokolsky}{Jee
  et~al\mbox{.}}{2010}]%
        {jee2010assurance}
\bibfield{author}{\bibinfo{person}{Eunkyoung Jee}, \bibinfo{person}{Insup Lee},
  {and} \bibinfo{person}{Oleg Sokolsky}.} \bibinfo{year}{2010}\natexlab{}.
\newblock \showarticletitle{Assurance cases in model-driven development of the
  pacemaker software}. In \bibinfo{booktitle}{{\em In Proceedings of the
  Leveraging Applications of Formal Methods, Verification, and Validation}}.
  \bibinfo{publisher}{Springer Berlin Heidelberg}, \bibinfo{pages}{343--356}.
\newblock


\bibitem[\protect\citeauthoryear{Kelly}{Kelly}{1999}]%
        {Kelly1999Arguing}
\bibfield{author}{\bibinfo{person}{Tim Kelly}.}
  \bibinfo{year}{1999}\natexlab{}.
\newblock {\em \bibinfo{title}{Arguing safety: a systematic approach to
  managing safety cases}}.
\newblock PhD thesis. \bibinfo{school}{University of York},
  \bibinfo{address}{Heslington, York, England}.
\newblock


\bibitem[\protect\citeauthoryear{Kelly}{Kelly}{2004}]%
        {kelly2004systematic}
\bibfield{author}{\bibinfo{person}{Tim Kelly}.}
  \bibinfo{year}{2004}\natexlab{}.
\newblock \showarticletitle{A systematic approach to safety case management}.
\newblock \bibinfo{journal}{{\em Journal of Passenger Cars: Electronic and
  Electrical Systems\/}} \bibinfo{volume}{113}, \bibinfo{number}{7}
  (\bibinfo{year}{2004}), \bibinfo{pages}{257--266}.
\newblock


\bibitem[\protect\citeauthoryear{Kelly}{Kelly}{2012}]%
        {kelly2012safety}
\bibfield{author}{\bibinfo{person}{Tim Kelly}.}
  \bibinfo{year}{2012}\natexlab{}.
\newblock \showarticletitle{Safety case use in the defence industry}.
\newblock In \bibinfo{booktitle}{{\em Supplements to: Using safety cases in
  industry and healthcare}}. \bibinfo{publisher}{The Health Foundation},
  \bibinfo{address}{London}, \bibinfo{pages}{19--23}.
\newblock


\bibitem[\protect\citeauthoryear{Kelly, Bate, McDermid, and Burns}{Kelly
  et~al\mbox{.}}{1997}]%
        {kelly1997building}
\bibfield{author}{\bibinfo{person}{Tim Kelly}, \bibinfo{person}{Iain Bate},
  \bibinfo{person}{John McDermid}, {and} \bibinfo{person}{Alan Burns}.}
  \bibinfo{year}{1997}\natexlab{}.
\newblock \showarticletitle{Building a preliminary safety case: An example from
  aerospace}. In \bibinfo{booktitle}{{\em In Proceedings of the Australian
  Workshop on Industrial Experience with Safety Critical Systems and
  Software}}. \bibinfo{publisher}{Not available}, \bibinfo{pages}{1--10}.
\newblock


\bibitem[\protect\citeauthoryear{Kelly and Weaver}{Kelly and Weaver}{2004}]%
        {Weaver2004The}
\bibfield{author}{\bibinfo{person}{Tim Kelly} {and} \bibinfo{person}{Rob
  Weaver}.} \bibinfo{year}{2004}\natexlab{}.
\newblock \showarticletitle{The goal structuring notation--a safety argument
  notation}. In \bibinfo{booktitle}{{\em In Proceedings of the Dependable
  Systems and Networks 2004 Workshop on Assurance Cases}}.
  \bibinfo{publisher}{Citeseer}.
\newblock


\bibitem[\protect\citeauthoryear{Klarlund and M{\o}ller}{Klarlund and
  M{\o}ller}{2001}]%
        {Klarlund2001Mona}
\bibfield{author}{\bibinfo{person}{Nils Klarlund} {and} \bibinfo{person}{Anders
  M{\o}ller}.} \bibinfo{year}{2001}\natexlab{}.
\newblock \bibinfo{booktitle}{{\em Mona version 1.4: User manual}}.
\newblock \bibinfo{publisher}{BRICS, Department of Computer Science, University
  of Aarhus Denmark}.
\newblock


\bibitem[\protect\citeauthoryear{Larson, Hatcliff, and Chalin}{Larson
  et~al\mbox{.}}{2013}]%
        {larson2013open}
\bibfield{author}{\bibinfo{person}{Brian Larson}, \bibinfo{person}{John
  Hatcliff}, {and} \bibinfo{person}{Patrice Chalin}.}
  \bibinfo{year}{2013}\natexlab{}.
\newblock \showarticletitle{Open source patient-controlled analgesic pump
  requirements documentation}. In \bibinfo{booktitle}{{\em In Proceedings of
  the 5th International Workshop on Software Engineering in Health Care}}.
  \bibinfo{publisher}{IEEE}, \bibinfo{pages}{28--34}.
\newblock


\bibitem[\protect\citeauthoryear{Leveson}{Leveson}{2011}]%
        {leveson2011use}
\bibfield{author}{\bibinfo{person}{Nancy Leveson}.}
  \bibinfo{year}{2011}\natexlab{}.
\newblock \bibinfo{booktitle}{{\em The Use of Safety Cases in Certification and
  Regulation}}.
\newblock \bibinfo{type}{{T}echnical {R}eport}.
  \bibinfo{institution}{Massachusetts Institute of Technology Engineering
  Systems Division}.
\newblock


\bibitem[\protect\citeauthoryear{Lewis}{Lewis}{2009}]%
        {Lewis2009Safety}
\bibfield{author}{\bibinfo{person}{Robert Lewis}.}
  \bibinfo{year}{2009}\natexlab{}.
\newblock \showarticletitle{Safety case development as an information modelling
  problem}.
\newblock In \bibinfo{booktitle}{{\em Safety-Critical Systems: Problems,
  Process and Practice}}. \bibinfo{publisher}{Springer},
  \bibinfo{pages}{183--193}.
\newblock


\bibitem[\protect\citeauthoryear{Luo, van~den Brand, Li, and Saberi}{Luo
  et~al\mbox{.}}{2017}]%
        {Luo2017A}
\bibfield{author}{\bibinfo{person}{Yaping Luo}, \bibinfo{person}{Mark van~den
  Brand}, \bibinfo{person}{Zhuoao Li}, {and} \bibinfo{person}{Arash~Khabbaz
  Saberi}.} \bibinfo{year}{2017}\natexlab{}.
\newblock \showarticletitle{A systematic approach and tool support for
  {GSN}-based safety case assessment}.
\newblock \bibinfo{journal}{{\em Journal of Systems Architecture\/}}
  \bibinfo{volume}{76} (\bibinfo{year}{2017}), \bibinfo{pages}{1--16}.
\newblock


\bibitem[\protect\citeauthoryear{Maksimov, Kokaly, and Chechik}{Maksimov
  et~al\mbox{.}}{2019}]%
        {Maksimov2019A}
\bibfield{author}{\bibinfo{person}{Mike Maksimov}, \bibinfo{person}{Sahar
  Kokaly}, {and} \bibinfo{person}{Marsha Chechik}.}
  \bibinfo{year}{2019}\natexlab{}.
\newblock \showarticletitle{A survey of tool-supported assurance case
  assessment techniques}.
\newblock \bibinfo{journal}{{\it Comput. Surveys}} \bibinfo{volume}{52},
  \bibinfo{number}{5} (\bibinfo{year}{2019}), \bibinfo{pages}{1--34}.
\newblock


\bibitem[\protect\citeauthoryear{Matsuno}{Matsuno}{2011}]%
        {Matsuno2011D}
\bibfield{author}{\bibinfo{person}{Yutaka Matsuno}.}
  \bibinfo{year}{2011}\natexlab{}.
\newblock \showarticletitle{D-case editor: A typed assurance case editor}.
\newblock \bibinfo{journal}{{\em University of Tokyo\/}}
  (\bibinfo{year}{2011}).
\newblock


\bibitem[\protect\citeauthoryear{Medhurst and Embrey}{Medhurst and
  Embrey}{2012}]%
        {medhurst2012safety}
\bibfield{author}{\bibinfo{person}{John Medhurst} {and} \bibinfo{person}{David
  Embrey}.} \bibinfo{year}{2012}\natexlab{}.
\newblock \showarticletitle{Safety case use in the railway industry}.
\newblock In \bibinfo{booktitle}{{\em Supplements to: Using safety cases in
  industry and healthcare}}. \bibinfo{publisher}{The Health Foundation},
  \bibinfo{address}{London}, \bibinfo{pages}{65--74}.
\newblock


\bibitem[\protect\citeauthoryear{Mendes, Hall, Matos, and Silvestre}{Mendes
  et~al\mbox{.}}{2014}]%
        {mendes2014reforming}
\bibfield{author}{\bibinfo{person}{Pietro Mendes}, \bibinfo{person}{Jeremy
  Hall}, \bibinfo{person}{Stelvia Matos}, {and} \bibinfo{person}{Bruno
  Silvestre}.} \bibinfo{year}{2014}\natexlab{}.
\newblock \showarticletitle{Reforming {B}razil's offshore oil and gas safety
  regulatory framework: Lessons from Norway, the United Kingdom and the United
  States}.
\newblock \bibinfo{journal}{{\em Energy Policy\/}}  \bibinfo{volume}{74}
  (\bibinfo{year}{2014}), \bibinfo{pages}{443--453}.
\newblock


\bibitem[\protect\citeauthoryear{Netkachova, Netkachov, and
  Bloomfield}{Netkachova et~al\mbox{.}}{2014}]%
        {Netkachova2014Tool}
\bibfield{author}{\bibinfo{person}{Kateryna Netkachova},
  \bibinfo{person}{Oleksandr Netkachov}, {and} \bibinfo{person}{Robin
  Bloomfield}.} \bibinfo{year}{2014}\natexlab{}.
\newblock \showarticletitle{Tool support for assurance case building blocks}.
  In \bibinfo{booktitle}{{\em International Conference on Computer Safety,
  Reliability, and Security}}. \bibinfo{publisher}{Springer},
  \bibinfo{pages}{62--71}.
\newblock


\bibitem[\protect\citeauthoryear{OpenAI}{OpenAI}{2023a}]%
        {OpenAI2023Create}
\bibfield{author}{\bibinfo{person}{OpenAI}.} \bibinfo{year}{2023}\natexlab{a}.
\newblock \bibinfo{title}{Create chat completion}.
\newblock   (\bibinfo{year}{2023}).
\newblock
\newblock
\shownote{https://platform.openai.com/docs/api-reference/chat.}


\bibitem[\protect\citeauthoryear{OpenAI}{OpenAI}{2023b}]%
        {OpenAI2023GPT35}
\bibfield{author}{\bibinfo{person}{OpenAI}.} \bibinfo{year}{2023}\natexlab{b}.
\newblock \bibinfo{title}{{GPT}-3.5 Documentation}.
\newblock   (\bibinfo{year}{2023}).
\newblock
\newblock
\shownote{https://platform.openai.com/docs/models/gpt-3-5.}


\bibitem[\protect\citeauthoryear{OpenAI}{OpenAI}{2023c}]%
        {OpenAI2023GPT4}
\bibfield{author}{\bibinfo{person}{OpenAI}.} \bibinfo{year}{2023}\natexlab{c}.
\newblock \bibinfo{title}{{GPT}-4 Documentation}.
\newblock   (\bibinfo{year}{2023}).
\newblock
\newblock
\shownote{https://platform.openai.com/docs/models/gpt-4.}


\bibitem[\protect\citeauthoryear{Palin and Habli}{Palin and Habli}{2010}]%
        {palin2010assurance}
\bibfield{author}{\bibinfo{person}{Robert Palin} {and} \bibinfo{person}{Ibrahim
  Habli}.} \bibinfo{year}{2010}\natexlab{}.
\newblock \showarticletitle{Assurance of automotive safety--a safety case
  approach}. In \bibinfo{booktitle}{{\em In Proceedings of the Computer Safety,
  Reliability, and Security}}. \bibinfo{publisher}{Springer Berlin Heidelberg},
  \bibinfo{pages}{82--96}.
\newblock


\bibitem[\protect\citeauthoryear{Resolute}{Resolute}{2016}]%
        {2016Resolute}
\bibfield{author}{\bibinfo{person}{Resolute}.} \bibinfo{year}{2016}\natexlab{}.
\newblock   (\bibinfo{year}{2016}).
\newblock
\newblock
\shownote{https://github.com/smaccm/smaccm/.}


\bibitem[\protect\citeauthoryear{Rinehart, Knight, and Rowanhill}{Rinehart
  et~al\mbox{.}}{2015}]%
        {Rinehart2015Current}
\bibfield{author}{\bibinfo{person}{David~J Rinehart}, \bibinfo{person}{John~C
  Knight}, {and} \bibinfo{person}{Jonathan Rowanhill}.}
  \bibinfo{year}{2015}\natexlab{}.
\newblock \bibinfo{booktitle}{{\em Current practices in constructing and
  evaluating assurance cases with applications to aviation}}.
\newblock \bibinfo{publisher}{National Aeronautics and Space Administration,
  Langley Research Center}.
\newblock


\bibitem[\protect\citeauthoryear{Rinehart, Knight, and Rowanhill}{Rinehart
  et~al\mbox{.}}{2017}]%
        {Rinehart2017Understanding}
\bibfield{author}{\bibinfo{person}{David~J Rinehart}, \bibinfo{person}{John~C
  Knight}, {and} \bibinfo{person}{Jonathan Rowanhill}.}
  \bibinfo{year}{2017}\natexlab{}.
\newblock \bibinfo{booktitle}{{\em Understanding What It Means for Assurance
  Cases to ``Work''}}.
\newblock \bibinfo{type}{{T}echnical {R}eport}.
\newblock


\bibitem[\protect\citeauthoryear{Rossi, Van~Beek, and Walsh}{Rossi
  et~al\mbox{.}}{2008}]%
        {Rossi2008Constraint}
\bibfield{author}{\bibinfo{person}{Francesca Rossi}, \bibinfo{person}{Peter
  Van~Beek}, {and} \bibinfo{person}{Toby Walsh}.}
  \bibinfo{year}{2008}\natexlab{}.
\newblock \showarticletitle{Constraint programming}.
\newblock \bibinfo{journal}{{\em Foundations of Artificial Intelligence\/}}
  \bibinfo{volume}{3} (\bibinfo{year}{2008}), \bibinfo{pages}{181--211}.
\newblock


\bibitem[\protect\citeauthoryear{Rushby, Xu, Rangarajan, and Weaver}{Rushby
  et~al\mbox{.}}{2015}]%
        {rushby2015understanding}
\bibfield{author}{\bibinfo{person}{John Rushby}, \bibinfo{person}{Xidong Xu},
  \bibinfo{person}{Murali Rangarajan}, {and} \bibinfo{person}{Thomas Weaver}.}
  \bibinfo{year}{2015}\natexlab{}.
\newblock \bibinfo{booktitle}{{\em Understanding and evaluating assurance
  cases}}.
\newblock \bibinfo{type}{{T}echnical {R}eport}. \bibinfo{institution}{NASA
  Langley Research Center}.
\newblock


\bibitem[\protect\citeauthoryear{Shahzad, Sheltami, Shakshuki, and
  Shaikh}{Shahzad et~al\mbox{.}}{2016}]%
        {Shahzad2016A}
\bibfield{author}{\bibinfo{person}{Farrukh Shahzad}, \bibinfo{person}{Tarek~R
  Sheltami}, \bibinfo{person}{Elhadi~M Shakshuki}, {and} \bibinfo{person}{Omar
  Shaikh}.} \bibinfo{year}{2016}\natexlab{}.
\newblock \showarticletitle{A review of latest web tools and libraries for
  state-of-the-art visualization}.
\newblock \bibinfo{journal}{{\em Procedia Computer Science\/}}
  \bibinfo{volume}{98} (\bibinfo{year}{2016}), \bibinfo{pages}{100--106}.
\newblock


\bibitem[\protect\citeauthoryear{Sklyar and Kharchenko}{Sklyar and
  Kharchenko}{2020}]%
        {sklyar2020assurance}
\bibfield{author}{\bibinfo{person}{Vladimir Sklyar} {and}
  \bibinfo{person}{Vyacheslav Kharchenko}.} \bibinfo{year}{2020}\natexlab{}.
\newblock \showarticletitle{Assurance case for safety and security
  implementation: a survey of applications}.
\newblock \bibinfo{journal}{{\em International Journal of Computing\/}}
  \bibinfo{volume}{19}, \bibinfo{number}{4} (\bibinfo{year}{2020}),
  \bibinfo{pages}{610--619}.
\newblock


\bibitem[\protect\citeauthoryear{Sujan, Habli, Kelly, Pozzi, and Johnson}{Sujan
  et~al\mbox{.}}{2016}]%
        {Sujan2016Should}
\bibfield{author}{\bibinfo{person}{Mark~A Sujan}, \bibinfo{person}{Ibrahim
  Habli}, \bibinfo{person}{Tim~P Kelly}, \bibinfo{person}{Simone Pozzi}, {and}
  \bibinfo{person}{Christopher~W Johnson}.} \bibinfo{year}{2016}\natexlab{}.
\newblock \showarticletitle{Should healthcare providers do safety cases?
  {L}essons from a cross-industry review of safety case practices}.
\newblock \bibinfo{journal}{{\em Safety Science\/}}  \bibinfo{volume}{84}
  (\bibinfo{year}{2016}), \bibinfo{pages}{181--189}.
\newblock


\bibitem[\protect\citeauthoryear{Toulmin}{Toulmin}{2003}]%
        {toulmin1958uses}
\bibfield{author}{\bibinfo{person}{Stephen Toulmin}.}
  \bibinfo{year}{2003}\natexlab{}.
\newblock \bibinfo{booktitle}{{\em The Uses of Argument}}.
\newblock \bibinfo{publisher}{Cambridge university press},
  \bibinfo{address}{England}. 1--247 pages.
\newblock


\bibitem[\protect\citeauthoryear{Vaswani, Shazeer, Parmar, Uszkoreit, Jones,
  Gomez, Kaiser, and Polosukhin}{Vaswani et~al\mbox{.}}{2017}]%
        {vaswani2017attention}
\bibfield{author}{\bibinfo{person}{Ashish Vaswani}, \bibinfo{person}{Noam
  Shazeer}, \bibinfo{person}{Niki Parmar}, \bibinfo{person}{Jakob Uszkoreit},
  \bibinfo{person}{Llion Jones}, \bibinfo{person}{Aidan~N Gomez},
  \bibinfo{person}{{\L}ukasz Kaiser}, {and} \bibinfo{person}{Illia
  Polosukhin}.} \bibinfo{year}{2017}\natexlab{}.
\newblock \showarticletitle{Attention is all you need}.
\newblock \bibinfo{journal}{{\em Advances in neural information processing
  systems\/}}  \bibinfo{volume}{30} (\bibinfo{year}{2017}).
\newblock


\bibitem[\protect\citeauthoryear{Vierhauser, Bayley, Wyngaard, Xiong, Cheng,
  Huseman, Lutz, and Cleland-Huang}{Vierhauser et~al\mbox{.}}{2019}]%
        {vierhauser2019interlocking}
\bibfield{author}{\bibinfo{person}{Michael Vierhauser}, \bibinfo{person}{Sean
  Bayley}, \bibinfo{person}{Jane Wyngaard}, \bibinfo{person}{Wandi Xiong},
  \bibinfo{person}{Jinghui Cheng}, \bibinfo{person}{Joshua Huseman},
  \bibinfo{person}{Robyn Lutz}, {and} \bibinfo{person}{Jane Cleland-Huang}.}
  \bibinfo{year}{2019}\natexlab{}.
\newblock \showarticletitle{Interlocking safety cases for unmanned autonomous
  systems in shared airspaces}.
\newblock \bibinfo{journal}{{\em IEEE transactions on software engineering\/}}
  \bibinfo{volume}{47}, \bibinfo{number}{5} (\bibinfo{year}{2019}),
  \bibinfo{pages}{899--918}.
\newblock


\bibitem[\protect\citeauthoryear{Voss, Sch{\"a}tz, Khalil, and Carlan}{Voss
  et~al\mbox{.}}{2013}]%
        {Voss2013Towards}
\bibfield{author}{\bibinfo{person}{Sebastian Voss}, \bibinfo{person}{Bernhard
  Sch{\"a}tz}, \bibinfo{person}{Maged Khalil}, {and} \bibinfo{person}{Carmen
  Carlan}.} \bibinfo{year}{2013}\natexlab{}.
\newblock \showarticletitle{Towards modular certification using integrated
  model-based safety cases}. In \bibinfo{booktitle}{{\em Proc. VeriSure:
  Verification and Assurance Workshop}}.
\newblock


\bibitem[\protect\citeauthoryear{Wassyng, Maibaum, Lawford, and Bherer}{Wassyng
  et~al\mbox{.}}{2011}]%
        {wassyng2011software}
\bibfield{author}{\bibinfo{person}{Alan Wassyng}, \bibinfo{person}{Tom
  Maibaum}, \bibinfo{person}{Mark Lawford}, {and} \bibinfo{person}{Hans
  Bherer}.} \bibinfo{year}{2011}\natexlab{}.
\newblock \showarticletitle{Software certification: Is there a case against
  safety cases?}. In \bibinfo{booktitle}{{\em In Proceedings of the Foundations
  of Computer Software. Modeling, Development, and Verification of Adaptive
  Systems}}. \bibinfo{publisher}{Springer Berlin Heidelberg},
  \bibinfo{pages}{206--227}.
\newblock


\bibitem[\protect\citeauthoryear{Widowati, Sutomo, and Istiono}{Widowati
  et~al\mbox{.}}{2021}]%
        {widowati2021elementary}
\bibfield{author}{\bibinfo{person}{Evi Widowati}, \bibinfo{person}{Adi Sutomo},
  {and} \bibinfo{person}{Wahyudi Istiono}.} \bibinfo{year}{2021}\natexlab{}.
\newblock \showarticletitle{Are Elementary Schools Ready for Disaster
  Preparedness and Safety?}
\newblock \bibinfo{journal}{{\em E3S Web Conf.\/}}  \bibinfo{volume}{317}
  (\bibinfo{year}{2021}), \bibinfo{pages}{1--13}.
\newblock


\bibitem[\protect\citeauthoryear{Willman}{Willman}{2021}]%
        {Willman2021Overview}
\bibfield{author}{\bibinfo{person}{Joshua Willman}.}
  \bibinfo{year}{2021}\natexlab{}.
\newblock \showarticletitle{Overview of {PyQt}5}.
\newblock In \bibinfo{booktitle}{{\em Modern PyQt}}.
  \bibinfo{publisher}{Springer}, \bibinfo{pages}{1--42}.
\newblock


\end{thebibliography}





\appendix

\section{Conversion between GSN and TDT Formats}\label{appendix:CubeSat_TDT_GSN}
\begin{landscape}
\begin{figure}[H]
\centering
\includegraphics[height=0.9\textheight]{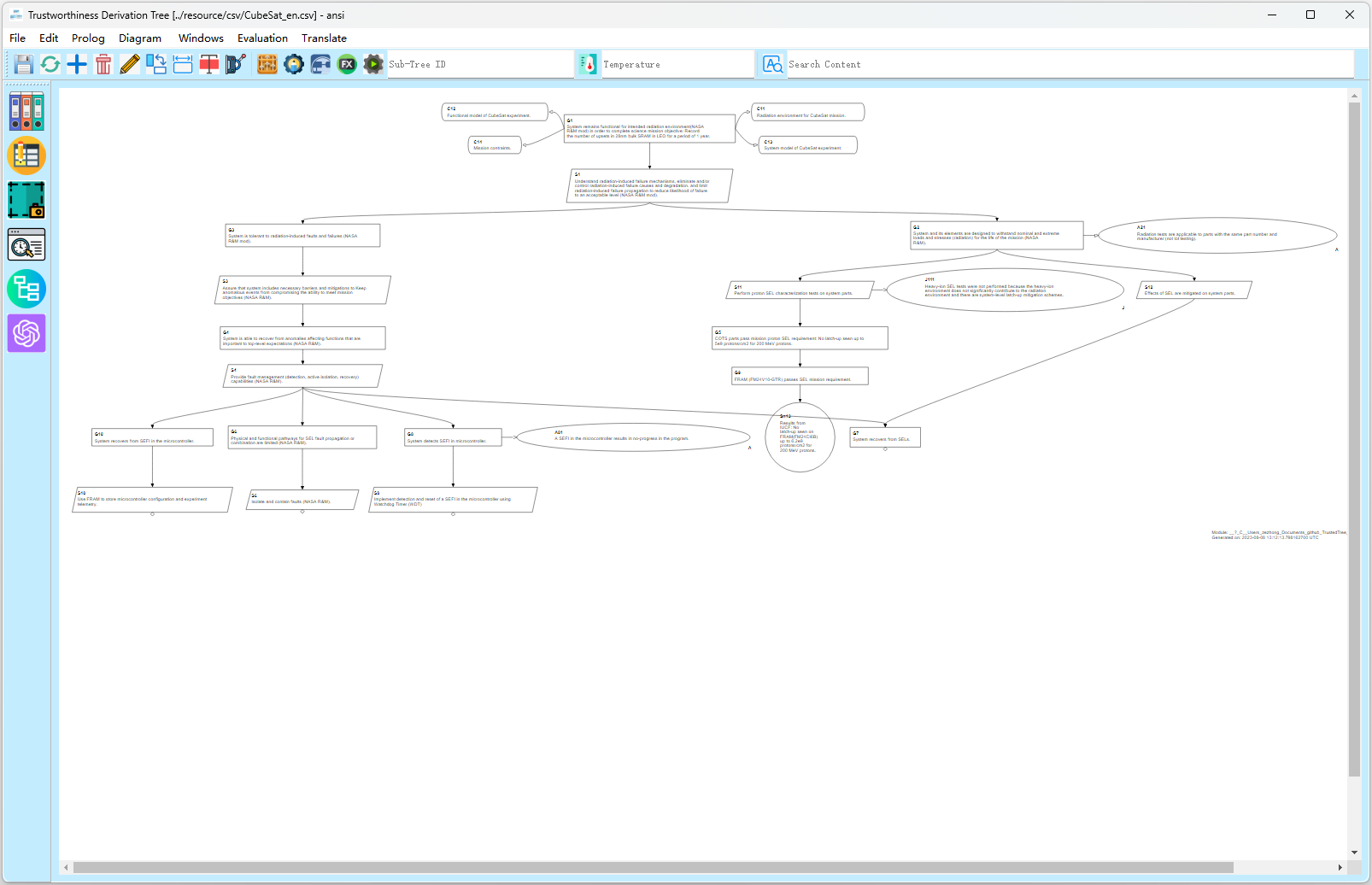}
\caption{A CubeSat-payload radiation-reliability assurance case using goal structuring notation.}
\label{ax_cubeset_gsn}
\end{figure}
\end{landscape}

\begin{landscape}
\begin{figure}[H]
\centering
\vskip 6.5cm
\includegraphics[width=1.4\textwidth]{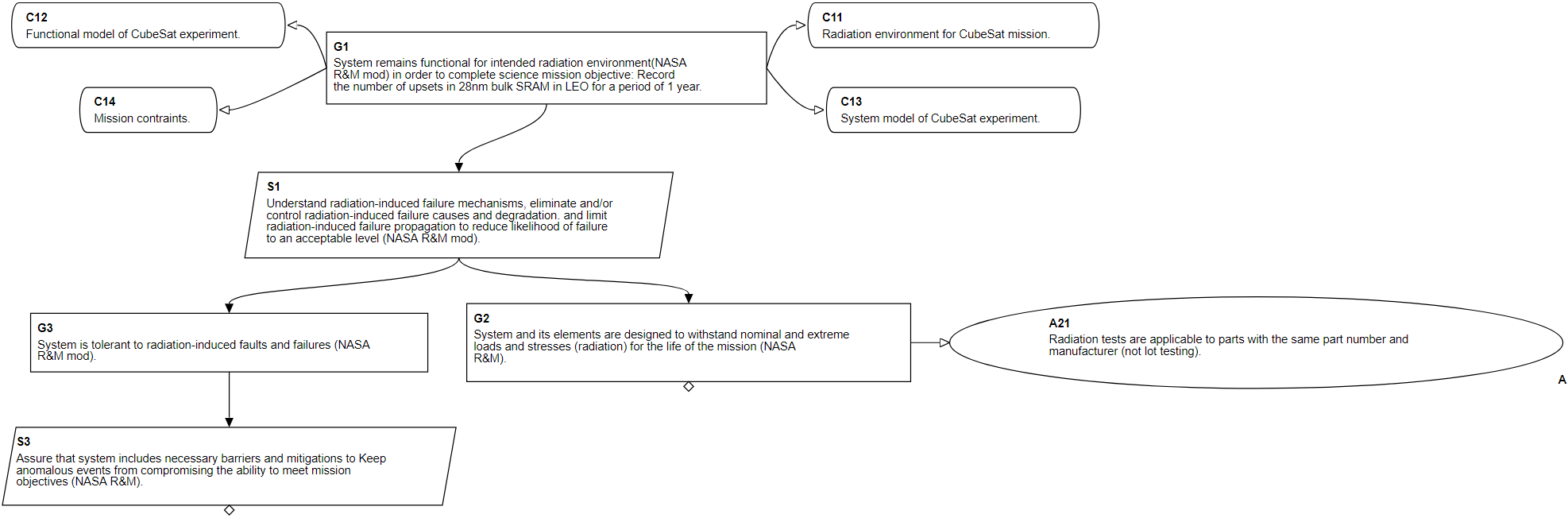}
\caption{Top-level GSN hierarchy.}
\label{ax_cubeset_gsn_d1}
\end{figure}
\end{landscape}

\begin{landscape}
\begin{figure}
\centering
\vskip 5cm
\includegraphics[width=1.45\textwidth]{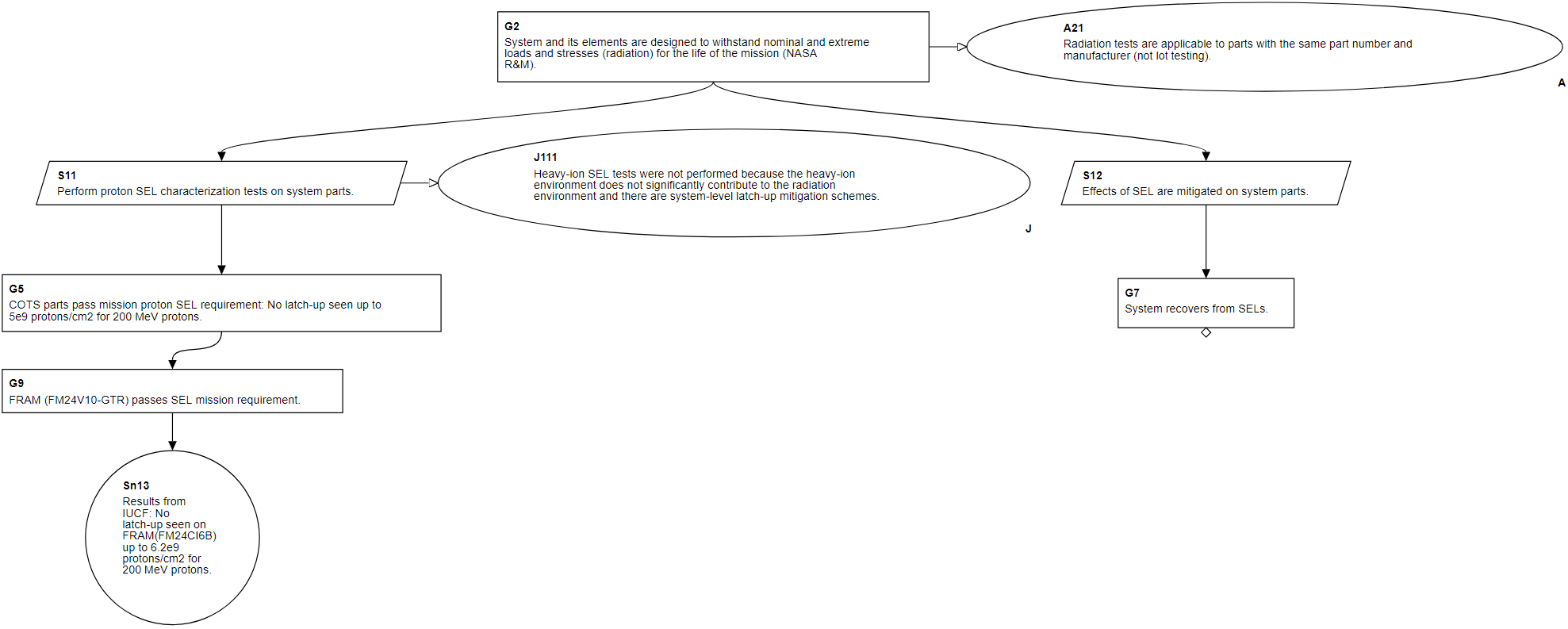}
\caption{Parts characterization hierarchy.}
\label{ax_cubeset_gsn_d2}
\end{figure}
\end{landscape}

\begin{landscape}
\begin{figure}
\centering
\vskip 6cm
\includegraphics[width=1.45\textwidth]{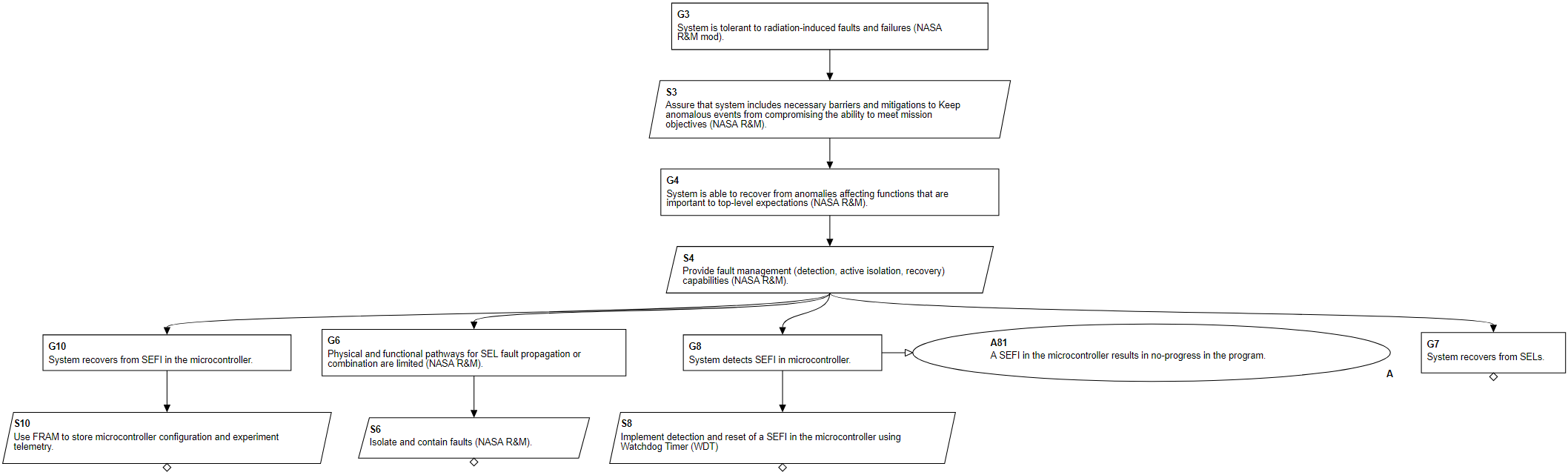}
\caption{System-level mitigation hierarchy.}
\label{ax_cubeset_gsn_d3}
\end{figure}
\end{landscape}

\begin{figure}
\centering
\includegraphics[width=1\textwidth]{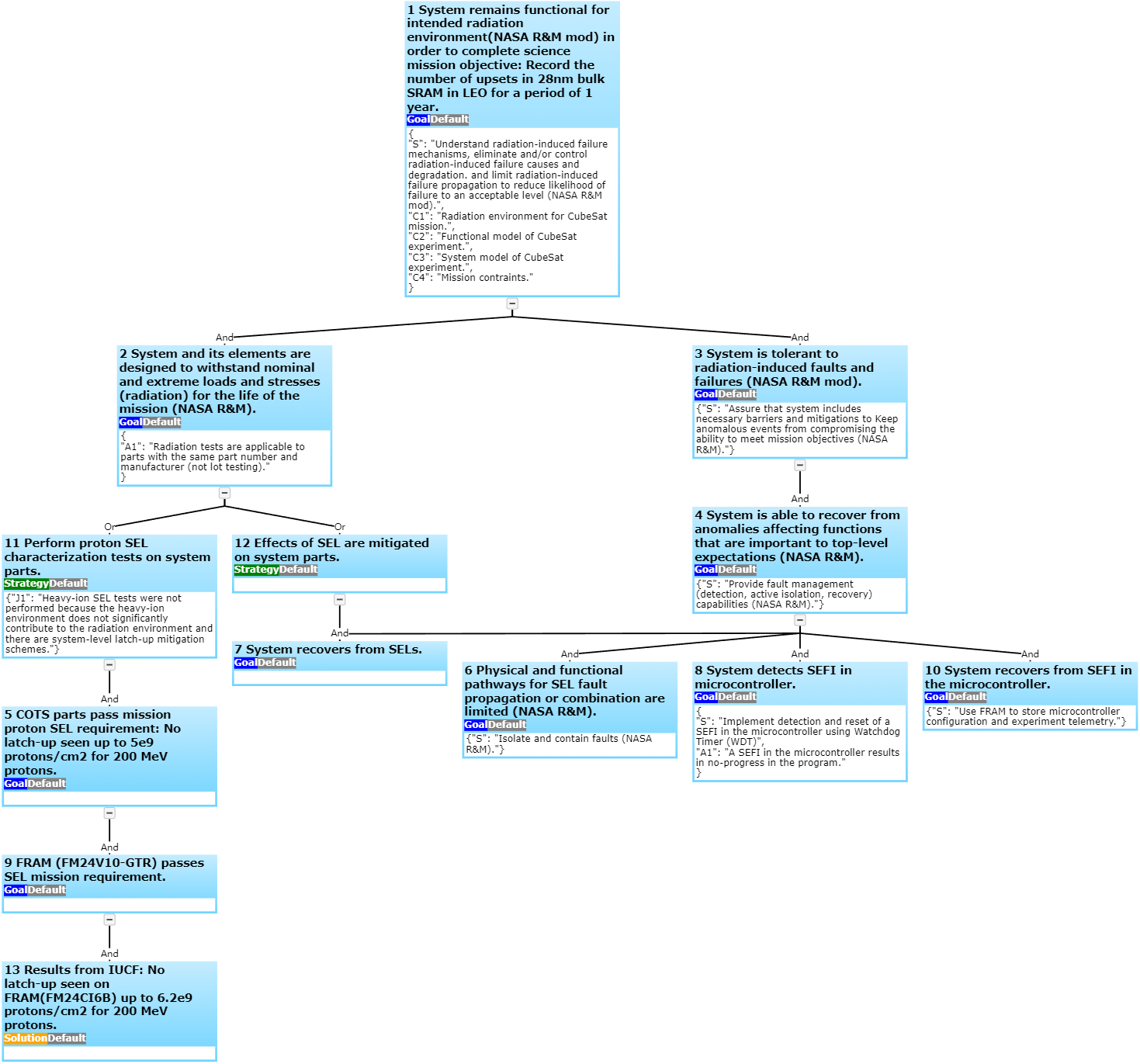}
\caption{A CubeSat-payload radiation-reliability assurance case using trustworthiness derivation tree.}
\label{ax_cubeset_tdt}
\end{figure}

\zz{
In this appendix, we present an illustrative example demonstrating the mutual conversion between an assurance case in Goal Structuring Notation (GSN) format and a Trustworthiness Derivation Tree (TDT). The example is inspired by the work of Austin et al.~\cite{austin2017cubesat}, where they employ GSN to express an assurance case for system-level mitigation of radiation effects in a CubeSat science experiment.

Figure~\ref{ax_cubeset_gsn} shows the original GSN, and Figures~\ref{ax_cubeset_gsn_d1}, \ref{ax_cubeset_gsn_d2}, and \ref{ax_cubeset_gsn_d3} are enlargements displaying various parts of Figure~\ref{ax_cubeset_gsn} in detail. Figure~\ref{ax_cubeset_tdt} represents the conversion into TDT format. In fact, TDT can also be translated back into GSN format using the Trusta tool. It can be observed that without losing any information, the TDT representation is more compact and emphasizes the key points, making it easier to read.
}

\zz{By providing the GSN to TDT conversion, we offer a bridge between traditional assurance case methodology and the more automated, formalized process enabled by Trusta. This facilitates a smooth transition for practitioners familiar with GSN, opening the door to the benefits of automatic reasoning and error detection in the assurance case development process.}

\section{The GSN format of the AGV example}\label{appendix:AGV_GSN}
\zz{
This appendix presents two figures. Figure~\ref{AGV_GSN_ALL_TDT} depicts the comprehensive TDT, serving as the complete version of what is shown in Figure~\ref{study_case_UI_after} from the main text. Figure~\ref{AGV_GSN_ALL} provides a graphical representation of this TDT in the GSN format.

In the TDT of Figure~\ref{AGV_GSN_ALL_TDT}, each node corresponds directly to either a goal or solution in the GSN representation. When transitioning to the GSN format, the auxiliary components, namely contexts, assumptions, justifications, and strategies, are captured within the descriptions of the TDT nodes in Figure~\ref{AGV_GSN_ALL_TDT}.

Collectively, these figures present a robust argument, underscoring the ability of the AGV to safely brake when encountering obstacles, thereby visualizing the detailed assurance case.
}

\newgeometry{left=1.5cm, right=1cm, top=1.5cm, bottom=1.5cm}
\begin{landscape}
\begin{figure}
\centering
\includegraphics[width=22cm, height=13cm, angle=0]{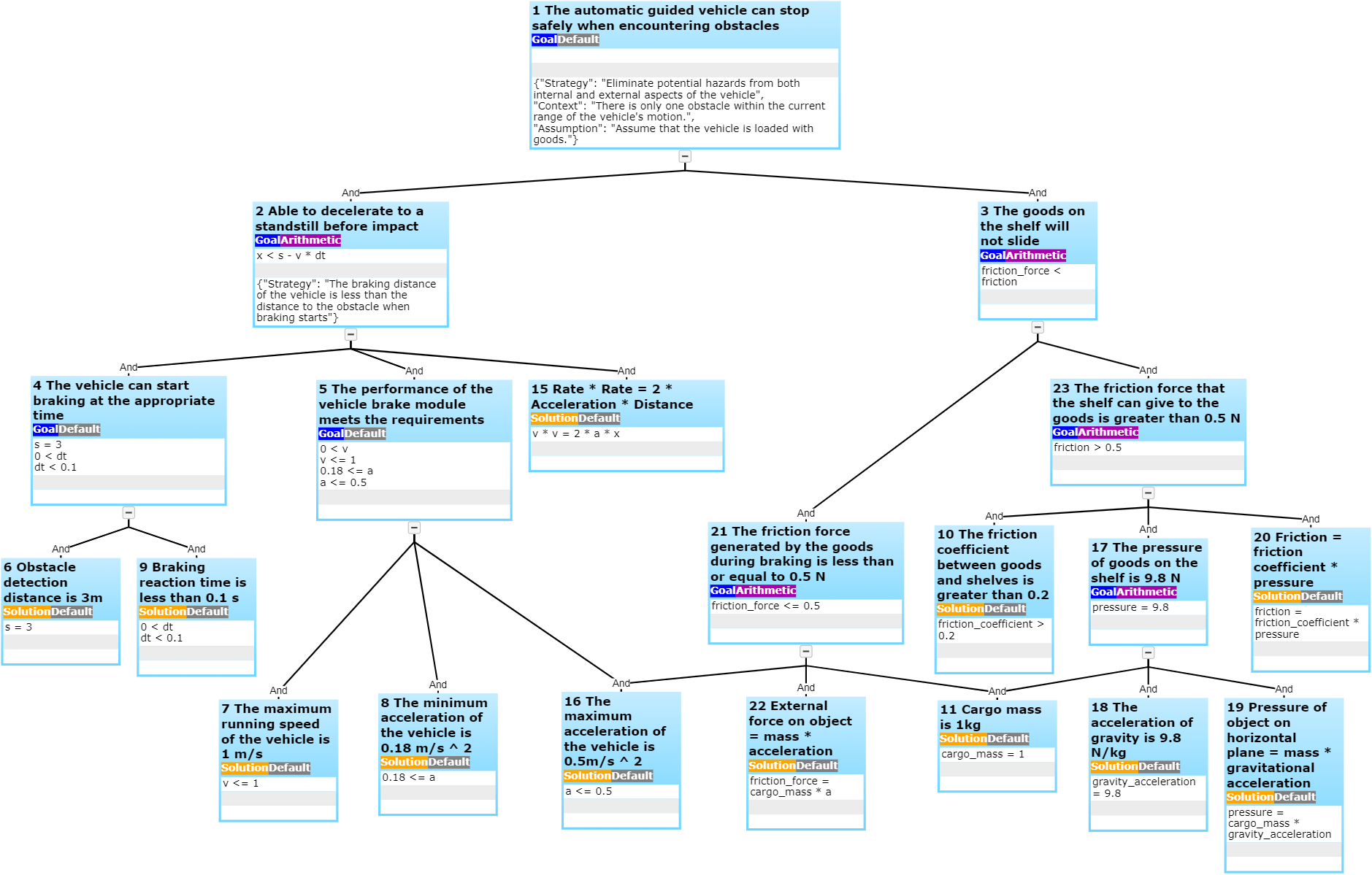}
\caption{The AGV example in TDT format.}
\label{AGV_GSN_ALL_TDT}
\end{figure}

\begin{figure}
\centering
\vskip 3cm
\includegraphics[width=22cm, height=10cm, angle=0]{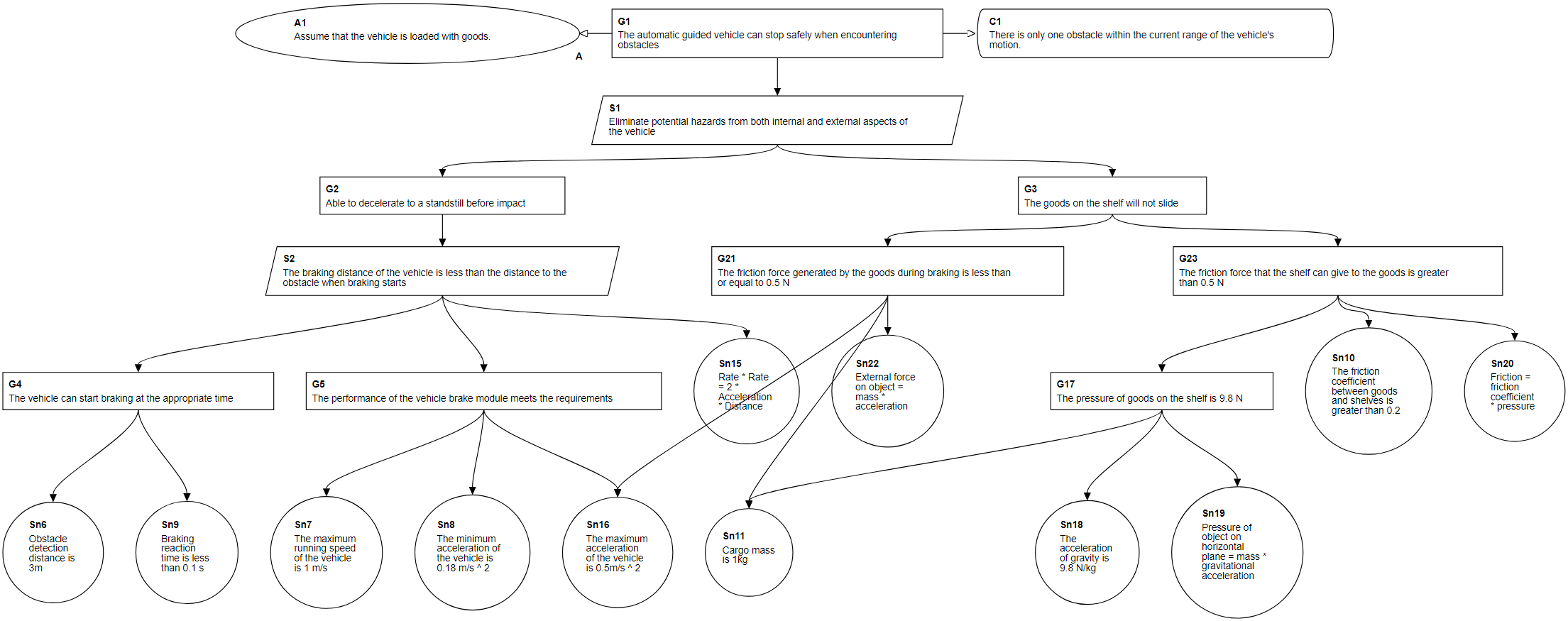}
\caption{The AGV example in GSN format.}
\label{AGV_GSN_ALL}
\end{figure}
\end{landscape}
\restoregeometry

\end{document}